\documentclass[%
 reprint,
 amsmath,amssymb,
 aps,
 longbibliography,
 pre,
floatfix,
nofootinbib,
]{revtex4-2}

\usepackage{graphicx}
\usepackage{dcolumn}
\usepackage{bm}
\usepackage{url}
\usepackage[ruled,vlined]{algorithm2e}
\usepackage{stmaryrd}
\DeclareMathOperator{\sign}{sign}

\usepackage{hyperref}
\renewcommand{\ref}[1]{%
  \hyperref[#1]{\textup{\tagform@{\ref*{#1}}}}%
}

\begin{document}

\title{Barriers and Dynamical Paths in Alternating Gibbs Sampling of \\ Restricted Boltzmann Machines}.

\author{Cl\'ement Roussel, Simona Cocco, R\'emi Monasson}

\address{Laboratory of Physics of the \'Ecole Normale Sup\'erieure, CNRS UMR 8023 $\&$ PSL Research, Sorbonne Universit\'e, 24 rue Lhomond, 75005 Paris, France}

\begin{abstract}
Restricted Boltzmann Machines (RBM) are bi-layer neural networks used for the unsupervised learning of model distributions from data. The bipartite architecture of RBM naturally defines an elegant sampling procedure, called Alternating Gibbs Sampling (AGS), where the configurations of the latent-variable layer are sampled conditional to the data-variable layer, and vice versa. We study here the performance of AGS on several analytically tractable models borrowed from statistical mechanics. We show that standard AGS is not more efficient than classical Metropolis-Hastings (MH) sampling of the effective energy landscape defined on the data layer. However, RBM can identify meaningful representations of training data in their latent space. Furthermore, using these representations and combining Gibbs sampling with the MH algorithm in the latent space can enhance the sampling performance of the RBM when the hidden units encode weakly dependent features of the data. 
We illustrate our findings on three datasets: Bars and Stripes and MNIST, well known in machine learning, and the so-called Lattice Proteins, introduced in theoretical biology to study the sequence-to-structure mapping in proteins. 
\end{abstract}

\maketitle

\section{Introduction}

Studying large heterogeneous and strongly interacting systems is a challenge common to various scientific fields. For decades, various numerical methods have been developed to sample high-dimensional configurations of such systems. Among these Monte-Carlo (MC) methods are one of the most powerful and versatile procedures \cite{metropolis_monte_1949,hastings_monte_1970}. Statistical averages over a target distribution are evaluated through an average over a set of stochastic configurations, generated according to a dynamical sampling process.
Nevertheless, it is a well-known issue that these methods can suffer from poor mixing: sampled configurations can be trapped in one of the regions of high probability, i.e., of low free energy, while other favorable regions are not dynamically explored. Therefore, it is of most importance to design sampling procedures capable of efficient exploration, allowing for fast transitions from one minimum of the free energy to another. For ferromagnetic systems, cluster algorithms, which identify and flip large clusters of spins at once achieve this objective 
\cite{swendsen_nonuniversal_1987,wolff_collective_1989,wang_cluster_1990,barbu_generalizing_2005}. 

\begin{figure*}[!htb]
\includegraphics{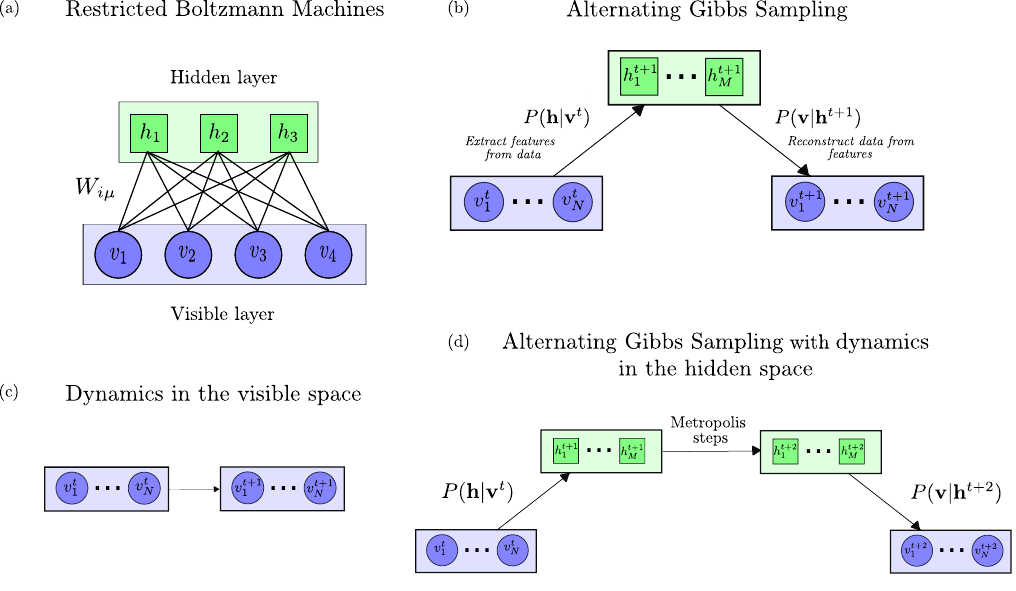}
\vskip -0.5cm
\caption{Description of the Restricted Boltzmann Machines and of the different sampling algorithms. (a) Bipartite architecture of RBM, with the visible (blue) and hidden (green) layers. (b) Alternating Gibbs Sampling: hidden and visible configurations are conditionally sampled from one another. (c) Sampling dynamics in the landscape $E^\text{eff}(\mathbf{v})$. (d) Modified Alternating Gibbs Sampling with dynamics in the hidden configuration space.}
\label{fig:algo}
\end{figure*}

Recently, machine learning algorithms have been developed to detect relevant MC updates in condensed matter models \cite{liu_self-learning_2017,xu_self-learning_2017,huang_accelerated_2017,nagai_self-learning_2017,shen_self-learning_2018,nagai_self-learning_2020}. Artificial neural networks are used to efficiently generate (with MC methods) low-energy configurations of approximate versions of target Hamiltonians. Hereafter we focus on one well-known machine learning architecture for unsupervised learning, called Restricted Boltzmann Machines (RBM) \cite{smolensky_chapter_1986,hinton_training_2002,tubiana_emergence_2017}. As illustrated in Fig.~\ref{fig:algo}(a), RBM are undirected graphical models constituted by two sets of interconnected random variables: a visible layer $\mathbf{v}$ that represents the data and a hidden layer $\mathbf{h}$ able to extract and explain their statistical features.
RBM learn a joint Boltzmann distribution $P(\mathbf{v},\mathbf{h})$ by maximizing the log-likelihood of the data configurations:
\begin{equation}
\label{eq:joint_distribution_v}
P(\mathbf{v}) = \int d\mathbf{h}\, P(\mathbf{v},\mathbf{h}) \ ,
\end{equation}
where the joint distribution of visible and hidden configurations reads
\begin{equation}
\label{eq:joint_distribution_v_h}
P(\mathbf{v},\mathbf{h}) = \frac 1{Z} \exp{\left(-E(\mathbf{v},\mathbf{h})\right)} \ ,
\end{equation}
and the energy $E(\mathbf{v},\mathbf{h})$ includes couplings between, but not within the layers. 
RBM have been widely studied from a statistical mechanics point of view \cite{agliari_multitasking_2012,tubiana_emergence_2017,decelle_spectral_2017,decelle_thermodynamics_2018,barra_phase_2018,hartnett_replica_2018}, see \cite{decelle_restricted_2020} for a recent review. 

The bipartite architecture of RBM suggests a natural procedure for sampling the marginal distribution $P(\mathbf{v})$. The method, called Alternating Gibbs Sampling (AGS), uses the conditional distributions $P(\mathbf{h}|\mathbf{v})$ and $P(\mathbf{v}|\mathbf{h})$ to sequentially sample the hidden and the visible spaces (Fig.~\ref{fig:algo}(b)). As the interaction graph is bipartite, the two conditional distributions factorize over the units of the sampled layer, which allows for independent draws of unit values (within a layer).

Despite its elegance and the simplicity of implementation, it is unclear whether AGS thermalizes substantially better than standard MC procedures in the effective energy landscape over the visible configurations, 
\begin{equation}
\label{eq:energy_v}
E^\text{eff}(\mathbf{v}) = - \log P(\mathbf{v}) \ ,
\end{equation}
see Fig.~\ref{fig:algo}(c). On the one hand, the conditional sampling of visible configurations through $P(\mathbf{v}|\mathbf{h})$ seems to allow for global moves in the $\mathbf{v}$ space, as with cluster algorithms. On the other hand, the conditional sampling of latent variables through $P(\mathbf{h}|\mathbf{v})$ indicates that their values reflect global features of visible configurations and could remain frozen when the system is stuck in free energy minima. The purpose of the present work is to investigate this question on a few analytically tractable models. We show that canonical AGS is generally not more efficient than naive Metropolis-Hastings algorithm in the visible landscape $E^\text{eff}(\mathbf{v})$. However, the architecture of RBM offers two advantages with respect to the latter. First, the sampling paths joining one free energy minimum to another can be more easily interpreted in terms of trajectory in the hidden space than in the visible space. Secondly, we proposed an augmented version of AGS, in which intermediate moves in the hidden space are carried out (Fig.~\ref{fig:algo}(d)). We show that this new sampling procedure yields much reduced thermalization times if the statistical features attached to the hidden units are decorrelated enough.

Our paper is organized as follows. First, we define RBM, its sampling algorithm, the Alternating Gibbs Sampling between the visible and hidden layers, \cite{geman_stochastic_1984, hinton_fast_2006, tieleman_training_2008} and the different datasets we use for numerical experiments in Section~\ref{sec:RBM_description_and_datasets}. Then, in Section~\ref{sec:alternating_Gibbs_sampling}, we introduce the models under consideration and study how AGS samples them. In Section~\ref{sec:dynamics_hidden_space}, we show how moving from one representation to another in the hidden space can help to sample. Finally, conclusions and perspectives are reported in Section~\ref{sec:fin}.

\section{Model and datasets}
\label{sec:RBM_description_and_datasets}
\subsection{Restricted Boltzmann Machines}

\label{sec:RBM_description}

Restricted Boltzmann Machines are undirected probabilistic graphical models with two layers. A visible layer $\mathbf{v}$, which represents the data, is connected to a hidden layer $\mathbf{h}$ through a weight matrix $W$ (Fig.~\ref{fig:algo}(a)). 

The visible layer includes $N$ units $v_i$, and the hidden layer M units $h_{\mu}$, which can take discrete or continuous values. The joint probability distribution of the visible configuration $\mathbf{v} = \{v_i\}_{i=1 \ldots N}$ and of the hidden configuration $\mathbf{h} = \{h_{\mu}\}_{\mu=1 \ldots M}$ is defined in Eq.~(\ref{eq:joint_distribution_v_h}). The energy $E(\mathbf{v},\mathbf{h})$ is equal to
\begin{equation}
E(\mathbf{v},\mathbf{h}) \!=\! - \sum \limits_{i=1}^N \sum \limits_{\mu=1}^M W_{i\mu} v_i h_{\mu}\! +\! \sum \limits_{\mu=1}^M \mathcal{U}_{\mu}(h_{\mu}) \!+ \!\sum \limits_{i=1}^N \mathcal{V}_i(v_i) \ . 
\label{eq:energy_RBM}
\end{equation}
In the formula above, $\mathcal{U}_{\mu}$ and $\mathcal{V}_i$ are potentials acting on, respectively, $ h_{\mu}$ and $v_i$. 

The effective energy over the visible configuration is obtained by marginalizing over the hidden units, see Eqs.~(\ref{eq:joint_distribution_v}) and (\ref{eq:energy_v}), up to an additive constant, with the result
\begin{equation}
E^\text{eff}(\mathbf{v}) = \sum \limits_{i=1}^N \mathcal{V}_i(v_i) - \sum \limits_{\mu=1}^M \Gamma_{\mu}(I_{\mu}(\mathbf{v})) \ ,\label{eq:energy_v_RBM}
\end{equation}
where $I_{\mu}(\mathbf{v}) = \sum \limits_{i=1}^N W_{i\mu} v_i$ is the input received by hidden unit $h_{\mu}$ and $\Gamma_{\mu}(I) = \log \left(\int dh \exp{\left(- \mathcal{U}_{\mu}(h) + hI \right)}\right)$ is the cumulative generative function associated with the potential $\mathcal{U}_{\mu}$. Parameters $\Theta \equiv \{W_{i\mu}, \mathcal{U}_{\mu}, \mathcal{V}_{i}\}$ modulate the energy landscape $E^\text{eff}(\mathbf{v})$. RBM, with binary visible and hidden units, are known to be universal approximators (i.e., can approximate any distribution over the visible variables) when the number $M$ of hidden units goes to infinity \cite{le_roux_representational_2008}.

If the set of parameters $\Theta$ is known, the RBM model distribution is fully defined. However, expected values over the distribution are generally not tractable, and are estimated through MC methods. Different algorithms, based on Alternating Gibbs Sampling between the visible and hidden layers, are used to generate samples from $P(\mathbf{v})$, such as Contrastive Divergence \cite{hinton_fast_2006}, or Persistent Contrastive Divergence \cite{tieleman_training_2008}. The pseudo-code of AGS is given in Algorithm~\ref{algo:ags} (Fig.~\ref{fig:algo}(b)). It is mainly composed of two steps:
\begin{itemize}
\item Starting from a visible configuration $\mathbf{v}^t$ at time $t$, a hidden configuration $\mathbf{h}^{t+1}$ is drawn from $P(\mathbf{h}|\mathbf{v}^t)$. This step can be seen as a stochastic feature extraction from the configuration $\mathbf{v}^t$.
\item A new visible configuration $\mathbf{v}^{t+1}$ is drawn from $P\left(\mathbf{v}|\mathbf{h}^{t+1}\right)$. This step can be seen as a stochastic reconstruction of $\mathbf{v}$ from the latent configuration $\mathbf{h}^{t+1}$.
\end{itemize}

Note that AGS or its variations are also used during the learning phase. For a given training set of $L$ samples, $\{\mathbf{v}^\ell\}_{\ell=1\ldots L}$, the parameters $\Theta$ are found by maximizing the log-likelihood of the data, $ \frac{1}{L} \sum \limits_{\ell=1}^L \log P(\mathbf{v}^\ell) \equiv
 \langle \log P(\mathbf{v}) \rangle_{\text{data}}$. The maximization is done by gradient ascent. The general expression for the gradients is
 \begin{eqnarray}
 \frac{\partial LL}{\partial \Theta} = - \left\langle \frac{\partial E^\text{eff}(\mathbf{v})}{\partial \Theta} \right\rangle_{\text{data}} +\left \langle \frac{\partial E^\text{eff}(\mathbf{v})}{\partial \Theta} \right\rangle_{\text{model}}\ ,
 \label{eq:gradients_LL}
 \end{eqnarray}
where $\langle.\rangle_{\text{data}}$ denotes the expected value over the data and $\langle.\rangle_{\text{model}}$ over the model. We see that estimating the gradient requires computing averages over the RBM distribution at every step of the training process.

\begin{algorithm}[!htb]
\label{algo:ags}
\SetAlgoLined
 Pick $\mathbf{v}^0$ in the training set\;
 \For{$t \in \llbracket0,T\rrbracket$}{
 $\mathbf{h}^{t+1} \sim P(\mathbf{h}|\mathbf{v}^t)$\;
 $\mathbf{v}^{t+1} \sim P\left(\mathbf{v}|\mathbf{h}^{t+1}\right)$\;
 }
\caption{Alternating Gibbs Sampling}
\end{algorithm}

\subsection{Datasets}

We use different datasets to illustrate our theoretical results. For all datasets, we train RBM using the learning algorithm of \cite{tubiana_emergence_2017}, available from \href{https://github.com/jertubiana/PGM}{https://github.com/jertubiana/PGM}. We then study how AGS or other sampling algorithms sample the RBM distribution.

\label{sec:dataset}

\subsubsection{Bars and Stripes}

Bars and Stripes (BAS) dataset \cite{mackay_information_2003} is made of $L \times L$ binary synthetic images which contain either exclusively bars or exclusively stripes. There are $2^{L+1} - 1$ possible configurations (Fig.~\ref{fig:dataset}(a)).

\subsubsection{MNIST}

MNIST dataset \cite{lecun_mnist_1998} is a large dataset of $28 \times 28$ pixel images of handwritten digits. We limit ourselves to zeros and ones (Fig.~\ref{fig:dataset}(b)), two graphically far digits. We use the binarized version of MNIST: each pixel is white or black.

\subsubsection{Lattice Protein}

\begin{figure}[!htb]
\includegraphics{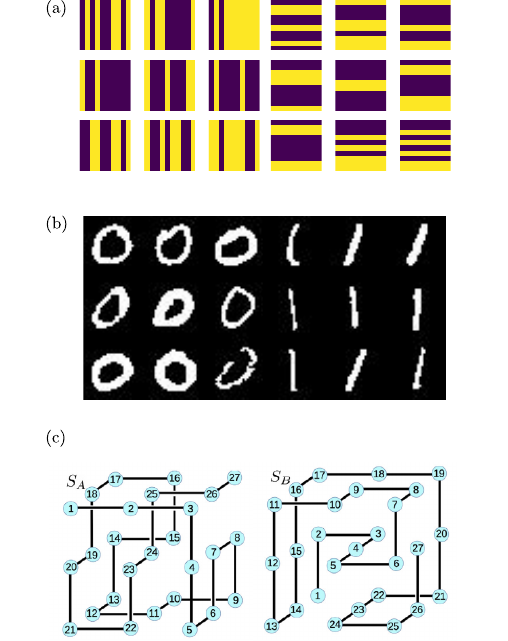}
\caption{(a) BAS: examples of bars (left) and stripes (right); here, $L = 10$. (b) MNIST: examples of hand-written 0 and 1 digits. (c) Lattice Proteins: two structures $S_A$ and $S_B$ defining two families of sequences having large $P_{\text{nat}}$ with either fold, see Eq.~(\ref{defpnat}). Structures from \cite{jacquin_benchmarking_2016}.}
\label{fig:dataset}
\end{figure}

Lattice Proteins (LP) are artificial proteins used to investigate protein design \cite{shakhnovich_enumeration_1990,mirny_protein_2001} and benchmarking inverse modeling procedures \cite{jacquin_benchmarking_2016}. Proteins are sequences of amino acids, whose 3D structures encode their functionalities. In this model, a structure is defined as a self-avoiding path of 27 amino-acid-long chains ($\mathbf{v}$ represents a sequence) on the $3 \times 3 \times 3$ lattice cube. There are  $\mathcal{N} = 103,406$ distinct structures (up to global symmetry). The probability that a protein sequence $\mathbf{v}$ folds in a given structure $S$ is given by
\begin{equation}\label{defpnat}
P_{\text{nat}}(S|\mathbf{v} ) = \frac{\exp{\left(-E(\mathbf{v},S) \right)} }{\sum \limits_{S' } \exp{\left(-E(\mathbf{v},S')\right)} } \ ,
\end{equation}
where the energy of the sequence $\mathbf{v}$ in a structure $S$ is defined through
\begin{equation}
E(\mathbf{v},S) = \sum \limits_{i<j} c^{S}_{i,j}\, E_{MJ} (v_i,v_j) \ .
\end{equation}
In the previous formula, $c^{S}_{i,j} = 1$ if the sites $i$ and $j$ are in contact (neighbors on the cube) in structure $S$; there are 28 contacts between the amino acids for each structure\footnote{Contacts along the chain are discarded, as their contribution to the energy is structure independent and, hence, does not affect the value of $P_{\text{nat}}$.}. Otherwise, $c^{S}_{i,j} = 0$. The pairwise energy $E_{MJ} (v_i,v_j)$ represents the physico-chemical interactions between the amino acids, given by the Miyazawa-Jernigan (MJ) potential \cite{miyazawa_residue-residue_1996}.
Here, we focus on two structures, $S_A$ and $S_B$, which define two protein families (Fig.~\ref{fig:dataset}(c)). For each structure, we sample $\sim 10^4$ sequences that have a high probability to fold in this structure ($P_{\text{nat}}(\mathbf{v} | S) > 0.99$) to build our datasets \cite{jacquin_benchmarking_2016}.

\section{Alternating Gibbs Sampling of multi-modal distributions}
\label{sec:alternating_Gibbs_sampling}

This section examines how long it takes for AGS to sample complex energy landscapes with several states associated with multi-modal distributions. We consider first the Curie-Weiss model at low temperature, where two ferromagnetic states with opposite magnetizations coexist. We then turn to the case of the Hopfield model, in which different, uncorrelated states coexist. We finally study the general, more complex situation, in which multiple correlated states are present, and the optimal sampling paths follow a well-defined ordering of the states.

\subsection{Case of bi-modal distribution}\label{sec:Curie_Weiss_model}

We consider the Curie-Weiss (CW) model over $N$ spins, $v_i=\pm \, 1$. The energy function is defined as 
\begin{equation}\label{defenerguCW}
E^{CW}(\mathbf{v}) = - \frac{w^2}{2N}\sum \limits_{i,j=1}^N  v_i v_j \ ,
\end{equation}
where $w^2$ plays the role of the inverse temperature. We start with the implementation of this mean-field model with RBM, before turning to a brief reminder of its properties and the study of the performance of AGS.

The CW model can be represented with a RBM with $N$ visible units (with potentials ${\cal V}_i=0$) and $M=1$ hidden unit with a quadratic potential $\mathcal{U}(h) = \frac{h^2}{2}$. The weights $W_{i,\mu=1}$ are uniform and equal to $\frac{w}{\sqrt{N}}$\footnote{We have checked that numerical experiments with RBM trained by gradient ascent on data sampled from the Curie-Weiss model converge to this solution.}. The energy of the RBM in Eq.~(\ref{eq:energy_RBM}) reads
\begin{equation}
\label{eq:RBM_Curie_Weiss_E(v,h)}
E^{CW}(\mathbf{v},\mathbf{h}) = - \frac{w}{\sqrt {N}} \sum \limits_{i=1}^N v_i \,h + \frac{h^2}{2} \ .
\end{equation}
After integration over $h$, it is straightforward to check that the effective energy in Eq.~(\ref{eq:energy_v_RBM}) coincides with the CW energy in Eq.~(\ref{defenerguCW}).

\subsubsection{Barriers and sampling time for MH procedures}

For $w^2>1$ and infinite-size limit $N \to \infty$, the average magnetization of the spins, $m = \frac{1}{N} \sum \limits_{i=1}^N v_i$, spontaneously acquires a non zero value. The value of this order parameter is determined by minimizing the free energy (per spin), $f(m) = -\frac{w^2 }{2} m^2 - \mathcal{S}(m),$ where 
\begin{equation}\label{defentbin}
    \mathcal{S}(m)= -\sum\limits_{\sigma = \pm \, 1} \frac{1+\sigma m }{2} \log \left(\frac{1+\sigma m }{2}\right) \ ,
\end{equation} 
is the entropy at fixed magnetization. The free energy $f(m)$ is an even function of $m$, with a double-well shape. The two opposite values of the spontaneous magnetization, roots of $f'(m^*)=0$, define two collective states of the system. Notice that $m=0$ is a local maximum of the free energy.

To go from one mode of the distribution to the other, a macroscopic number of spins has to be flipped. Local sampling processes, such as Metropolis-Hastings described in Algorithm~\ref{algo:MH}\footnote{The specific choice of the Metropolis rule is irrelevant here; other choices, such as Glauber rule, \cite{glauber_timedependent_1963}, do not affect the leading behavior of $\tau$.} take exponential-in-$N$ time to do so:
\begin{equation}\label{feb}
\tau \sim \exp{\left(N\Delta f\right)}\ , \ \ \ \text{where} \ \ \ \Delta f \equiv f(\pm \, m^*) - f(0) \ ,
\end{equation}
is the free energy barrier between the minima $m = \pm \, m^*$ and the local maximum $m = 0$ of the free energy landscape. Consequently, for large $N$, the system is stuck in one state/mode for long times, and thermalization is practically impossible.

\begin{algorithm}[!htb]
\SetAlgoLined
 Pick $\mathbf{v}^0 \in \{-1,1\}^N$ at random \;
 \For{$t \in \llbracket0,T\rrbracket$}{
 $\mathbf{v}^{'} = \mathbf{v}^t$ \;
 Choose $i \in \llbracket1,N\rrbracket$ uniformly at random\;
 $v'_i = - v_i^t$\;
 Generate a uniform random $u \in [0,1]$ \;
\uIf{$u \leq \min\left(1,\exp\big[-(E^{\text{eff}}(\mathbf{v}')-E^{\text{eff}}(\mathbf{v}^t))\big]\right)$}{
    \qquad $\mathbf{v}^{t+1} =\mathbf{v}'$ \;
  }
  \Else{
   \qquad $\mathbf{v}^{t+1}=\mathbf{v}^{t}$ \;
 }
 }
 \caption{Metropolis-Hastings algorithm}
 \label{algo:MH}
\end{algorithm}

\subsubsection{Optimal sampling paths with AGS}

The AGS procedure can be entirely described in terms of the magnetizations $m$ of the visible configurations and of the values $h$ of the hidden unit. To get intensive quantities in the large $N$ limit, we rescale $h\to h/\sqrt N$. The conditional configuration of the hidden unit $h^{t+1}$ given a visible configuration with magnetization $m^t$ then simply reads
\begin{equation}\label{eq0k}
P\big( h^{t+1}|m^t \big) = \frac 1{\sqrt{2\pi/N}}\, \exp\left( -\frac N2 \big( h^{t+1}-w\, m^t\big)^2\right) \ .
\end{equation}
Some care must be taken to write the conditional distribution of the magnetization $m^t$ given the hidden unit $h^t$. First, the conditional probability of $\mathbf{v}^t$ is
\begin{eqnarray}\label{eq1k}
P\big( \mathbf{v}^{t}|h^t \big) &=&\prod_{i=1}^N \frac{\exp{\left(w\,h^t\,v^t_i\right)}}{2\cosh (w\,h^t)} \\ &=& \exp{\left(N\big(w\,h^t\,m^t-\log 2\cosh(w\,h^t) \big)\right)} \nonumber \ ,
\end{eqnarray} 
which depends on $m^t$ as expected. Second, to turn the probability over visible configurations into a probability over magnetizations, we have to take into account the entropies of the latters. We end up with
the normalized (to dominant order in $N$) conditional probability
\begin{eqnarray}\label{eq2k}
P\big( m^{t}|h^t \big) &=& \exp{\left(N\big(w\,h^t\,m^t-\log 2\cosh(w\,h^t)\big)\right)} \nonumber \\
&\times& \exp{\left(N \, {\cal S}(m^t) \right)}  \ .
\end{eqnarray} 
We may now express the probability to go from one minimum of the free energy landscape to the other in $T$ steps of AGS. To do so, we compute the probability $P\left(m^T|m^0\right)$ that, given magnetization $m^0=m^*$ at time $t=0$, the dynamics associated with AGS reaches magnetization $m^T=-m^*$ at time $t=T$. This conditional probability may be computed by means of the saddle-point method in the thermodynamic limit $N \to \infty$ (for finite $T$):
\begin{widetext}
\begin{equation} 
 P\left(m^T|m^0\right) = \int dh^1  \ldots dh^{T} \int dm^1  \ldots  dm^{T-1} \nonumber \prod \limits_{t=0}^{T-1} P(m^{t+1}|h^{t+1})\,P(h^{t+1}|m^{t}) = \exp{\left(- N \min\limits_{\{m^t,h^t\}}\Phi\left(\{m^t,h^t\}\right)\right)}\ ,
\label{eq:integral_dynamics}
\end{equation}
\end{widetext}
where 
\begin{equation}
 \Phi\left(\{m^t,h^t\}\right) = \sum\limits_{t=0}^{T-1} \delta \Phi\left(t\to t+1\right) \ ,
 \end{equation}
 and, according to Eqs.~(\ref{eq0k}) and (\ref{eq2k}),
 \begin{eqnarray} \label{eqdeltaphi}
 \delta \Phi\left(t\to t+1\right) &=& \frac{1}{2}(h^{t+1}-w\, m^t)^2  \nonumber \\
 &+&\log \left(2 \cosh\left(w\, h^{t+1}\right)\right) \nonumber \\
 &-&w \,m^{t+1}\, h^{t+1} - \mathcal{S}\big(m^{t+1}\big)  \ .
\end{eqnarray}

The set of magnetizations $m^t$ and hidden-unit values $h^t$ minimizing the action $\Phi$ in Eq.~(\ref{eq:integral_dynamics}) define the most likely path, with AGS, capable of moving the system from one state to another in $T$ alternating sampling steps. They are solutions of the following extremization equations for $\Phi$, which must be fulfilled at all steps $1\le t\le T-1$:
\begin{eqnarray}\label{eqmotion1}
w(m^{t+1} + m^{t})\ & =& h^{t+1} +w \, \tanh(w\, h^{t+1}) \ , \\
\text{arctanh}\ (m^{t}) &=&  w (h^{t} +h^{t+1})\!-\!w^2 m^{t} \ .\nonumber
\end{eqnarray}

\begin{figure}[!htb]
\includegraphics{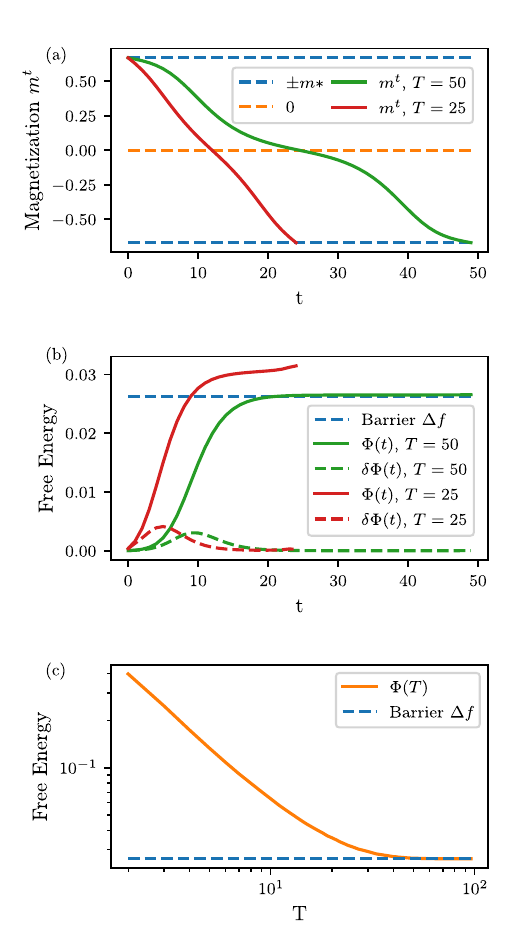}
\caption{Numerical minimization of $\Phi\left(\{m^t,h^t\}\right)$ for $w=1.1$ with boundary conditions $m^0=-m^T=m^*$. (a) Optimal time course of the magnetizations for $T=25$ (red) and $T=50$ (green) AGS steps. 
(b) Contributions $\delta \Phi(t)$ and full action $\Phi(t)$ as a function of the number of AGS steps for the optimal paths of duration $T=25$ and $T=50$. 
(c) Cost $\Phi$ of the optimal path as a function of $T$. For large $T$, $\Phi$ reaches from above a plateau equals to the free energy barrier $\Delta f$ of the CW model, see Eq.~(\ref{feb}). The convergence is exponentially fast, with decay time $T_\text{decay} \sim 1/\log(w^2)$.}
\label{fig:Curie-Weiss}
\end{figure}

An example of transition path obtained through brute force numerical minimization of $ \Phi\left(\{m^t,h^t\}\right)$ is shown in Fig.~\ref{fig:Curie-Weiss}(a). It is composed of two portions: 
\begin{itemize}
\item an initial part of the trajectory ascending the free energy landscape from one stable state, say, $+m^*$ up to the free energy local maximum, $m=0$. This part is associated with an exponentially small probability, i.e., to a positive contribution to the action, $\delta \Phi>0$ (Fig.~\ref{fig:Curie-Weiss}(b)). 
\item a final part of the trajectory descending the free energy landscape from the local maximum $m=0$ down to the other stable state, say, $-m^*$. This stretch does not seem to contribute to the action, $\delta \Phi\simeq 0$ (Fig.~\ref{fig:Curie-Weiss}(b)). 
\end{itemize}
As the number $T$ of steps increases the total action decreases, as expected, and quickly converges towards a minimal value (Fig.~\ref{fig:Curie-Weiss}(c)).
We show below that the scenario above can be analytically understood when $T$ is sent to infinity. 

\subsubsection{Analytical expressions of the optimal trajectories in the \texorpdfstring{$T\to\infty$}{T to infinity}  limit}

In the infinite $T$ limit, the equations of motion (\ref{eqmotion1}) admit two distinct solutions that correspond to the two-fold behavior empirically observed for finite $T$.

\paragraph{Instanton-like trajectories.}
The ascending trajectories correspond to instantons, connecting a local minimum of the free energy to the local maximum, and are described by 
\begin{eqnarray} m^{t+1} &=& \frac 1{w^2}\, \text{arctanh}\ (m^{t}) \ , \nonumber \\ \label{eq:instantonic_CW_1}
h^{t+1} &=& w\, m^{t+1} \ .  
\end{eqnarray}

Inserting these equations into Eq.~(\ref{eqdeltaphi}), the contribution to the action associated to one AGS step reads, after some algebra,
\begin{equation}
\delta \Phi = f\big( m^{t+1}\big) -f\big( m^t\big) \ , 
\end{equation}
where $f(m)$ is the free energy of the CW model for magnetization $m$.
The only stable fixed point of this dynamics is the local maximum of $f(m)$ in $m=0$. Starting from $m^0=m^*$, the dynamics converges to $m=0$ for $T \to \infty$. Along this path, $\Phi\left(\{m^t,h^t\}\right) \underset{T \to \infty}{\longrightarrow} f(0)-f(m^*)=\Delta f$ (Fig.~\ref{fig:Curie-Weiss}(c)). Hence this path has a log-probability (per variable) equal to minus the free energy barrier separating the minima of the landscape.

\paragraph{Thermalization-like trajectories.} The descending portion of the trajectory corresponds to relaxation towards the other minimum of the free energy and is described by the following solution of the extremization equations:
\begin{eqnarray}
m^{t+1} &=& \tanh \left(w^2\, m^{t}\right) \ ,\nonumber \\
\label{eq:relaxation_CW_2}
h^{t+1} &=& w\, m^{t}   \ .
\end{eqnarray}
We find that the contribution of an alternating step of AGS to the action vanishes
\begin{equation}
    \delta \Phi = 0 \ .
\end{equation}
The stable fixed points of the dynamics are the two minima of $f(m)$. Starting from $m^0=0$ at time $t=0$, the dynamics converges, when $T \to \infty$, to the spontaneous magnetization $\pm \, m^*$ associated to the minima of $f(m)$. Along this relaxation part of the trajectory, $\Phi\left(\{m^t,h^t\}\right) = 0$.

As a summary, the probability that a sequence of $T$ steps of Alternating Gibbs Sampling brings the system from one minimum of the free energy to the other is given, to the dominant order in $N$, by $\exp{(-N \Delta f)}$. This result holds when $N$ and $T$ are very large (but with $T\ll N$). We conclude that it will take the same time $\tau$ as with the MH procedure, see Eq.~(\ref{feb}), for the system to switch state. In other words, AGS is as inefficient as MH for sampling the bi-modal distribution associated with the CW model.

\subsection{Case of unstructured multi-modal distribution}\label{sec:hop}

We now consider the case of a multi-modal distribution, where more than two states have high probabilities. 

\subsubsection{Hopfield model}

Let us call $\boldsymbol{\xi}^\mu$ ($\mu=1...M$) the centers of the states, which we suppose to be orthogonal in the infinite $N$ limit. We assume that $\xi_i^\mu=\pm \, 1$. The order parameter is the $M$-dimensional vector of magnetizations along the centers, called patterns,
\begin{equation}\label{defvi4}
    m_\mu = \frac 1N \sum \limits_{i=1}^N \langle v_i\rangle \xi_i^\mu \ .
\end{equation}
We will hereafter consider the limit $\frac{M}{N} \to 0$. To be more precise, the energy over the visible configurations corresponds to the Hopfield model \cite{hopfield_neural_1982}, and is defined through 
\begin{equation}\label{defenergyHopfield}
E^{\text{Hop}}(\mathbf{v}) = - \frac{w^2}{2N}\sum \limits_{i,j=1}^N  \left(\sum \limits_{\mu=1}^M \xi^{\mu}_i \xi^{\mu}_j \right) v_i v_j \ ,
\end{equation}
where $w^2$ is the inverse temperature of the model, and the visible variables take values $v_i = \pm 1$. By inserting Eq.~(\ref{defvi4}) into Eq.~(\ref{defenergyHopfield}), the free energy (per variable) can be written as a function of the magnetizations $\mathbf{m}$ along the centers 
\begin{equation}
    \label{eq:fm_hopfield}
    f(\mathbf{m}) = -\frac{w^2}{2} \sum \limits_{\mu=1}^M m_{\mu}^2 - \mathcal{S}^{\text{Hop}}(\mathbf{m}) \ ,
\end{equation}
where $\mathcal{S}^{\text{Hop}}(\mathbf{m})$ denotes the entropy of the visible configurations at fixed magnetizations. It can be computed from the following Legendre formula
\begin{eqnarray}
 \mathcal{S}^{\text{Hop}}(\mathbf{m}) =\underset{\boldsymbol{\lambda}}{\text{min}}  \Bigg( &&\frac{1}{N} \sum \limits_{i=1}^N\log 2 \cosh \left(\sum \limits_{\mu=1}^M\xi_i^{\mu} \lambda_\mu  \right) \nonumber \\
 &&- \sum \limits_{\mu=1}^M \lambda_{\mu} m_{\mu} \Bigg) \ .
\end{eqnarray}
The minimum is reached in the unique $\boldsymbol{\lambda}^*$ such that
\begin{equation}
 m_\mu = \frac{1}{N} \sum _i  \xi_i^{\mu} \, \tanh  \left(\sum _\nu\xi_i^{\nu} \lambda^*_\nu  \right) \ ,
\end{equation}
for all $\mu$'s. $\mathcal{S}^{\text{Hop}}(\mathbf{m})$ can be expressed as a function of $\boldsymbol{\lambda}^*$ and the binary entropy $\mathcal{S}(m)$ defined in Eq.~(\ref{defentbin})
\begin{eqnarray}
 \mathcal{S}^{\text{Hop}}(\mathbf{m}) =  \frac{1}{N} \sum _i \mathcal{S}\left( \tanh \left(\sum _\mu \xi_i^{\mu} \lambda^*_\mu \right) \right)  \ .
 \end{eqnarray}

The Hopfield model can be represented with a RBM with $N$ visible units (with potentials ${\cal V}_i=0$) and $M$ hidden units subject to the quadratic potential $\mathcal{U}(h) = \frac{h^2}{2}$ \cite{barra_equivalence_2012,agliari_multitasking_2012,leonelli_effective_2021}. The energy of the RBM in Eq.~(\ref{eq:energy_RBM}) reads
\begin{equation}
\label{eq:RBM_Hopfield_E(v,h)}
E^{Hop}(\mathbf{v},\mathbf{h}) = -  \sum _{i,\mu} W_{i\mu} v_i \,h_{\mu} +  \sum _\mu \frac{h_{\mu}^2}{2}  \ .
\end{equation}
It is straightforward to check, after integration over the $M$ hidden units, that the effective energy in Eq.~(\ref{eq:energy_v_RBM}) coincides with the Hopfield energy in Eq.~(\ref{defenergyHopfield}) provided 
the weights fulfill the constraints
\begin{equation}\label{eq:weights_Hebb}
    \sum _{\mu} W_{i\mu} \, W_{j\mu} = \frac{w^2}{N} \sum \limits_\mu \xi^{\mu}_i \xi^{\mu}_j\ .
\end{equation}
These conditions do not uniquely define the weight matrix $\mathbf{W}$. The energy is invariant under any transformation $\mathbf{W} \rightarrow \mathbf{W}\times \mathbf{O}$, where $\mathbf{O}$ is an orthogonal matrix. We choose for now the following parametrization for the weight matrix $\mathbf{W}$:
\begin{equation}
\label{eq:weights_Hopfield}
    W_{i\mu} = \frac{w}{\sqrt N}\, \xi^{\mu}_i \ .
\end{equation}
Alternative choices will be discussed later.

\begin{figure*}[!htb]
\includegraphics{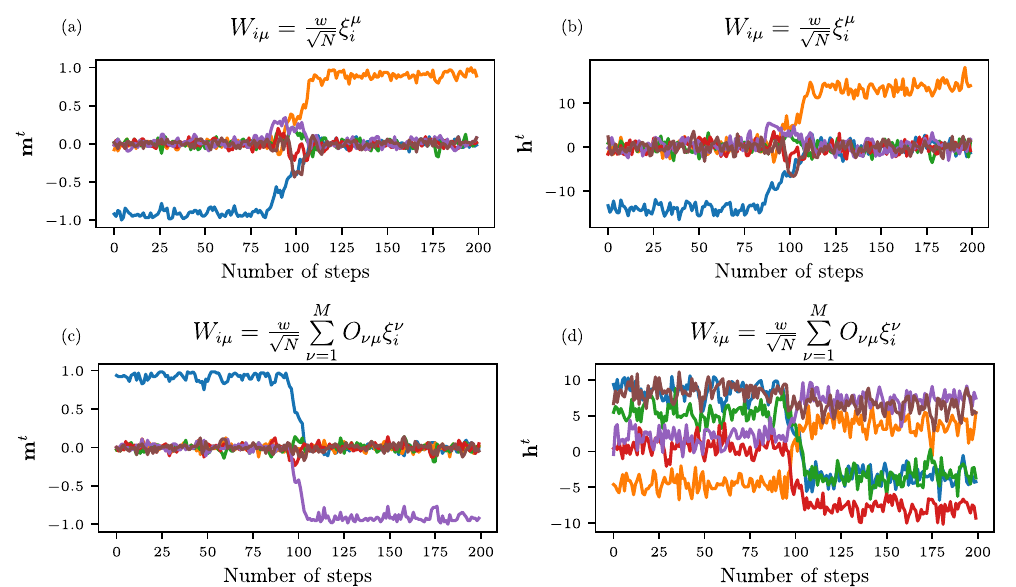}
\caption{Hopfield model with $N = 128$ spins, $M = 6$ patterns and $w = 1.35$; each color refers to one index $\mu$. Examples of transition between two states for $W_{i\mu} = \frac{w}{\sqrt{N}} \xi^{\mu}_{i}$ (panels a-b) and for $W_{i\mu} = \frac{w}{\sqrt{N}} \sum _\nu O_{\nu\mu} \xi^{\nu}_{i}$ (panels c-d). (a-c) Magnetizations $m_\mu$ along the patterns as functions of the number of AGS steps. (b-d) 
Hidden unit values $h_\mu$ as functions of the number of AGS steps for the same transitions as in panels (a-c). }
\label{fig:transition_Hopfield}
\end{figure*}

\subsubsection{Optimal sampling with AGS}

The AGS procedure can be entirely described in terms of $M$ magnetizations $\mathbf{m}$ of the visible configurations and of the values $\mathbf{h}$ of the M hidden units. As in the case of the CW model, to get intensive quantities in the large $N$ limit, we rescale $\mathbf{h}\to \mathbf{h}/\sqrt N$. The conditional configuration of the hidden unit $\mathbf{h}^{t+1}$ given a visible configuration with magnetization $\mathbf{m}^t$ is factorized, and reads
\begin{equation}
P\big( h_{\mu}^{t+1}|\mathbf{m}^t \big) = \frac 1{\sqrt{2\pi/N}}\, \exp\left( -\frac N2 \Big( h_{\mu}^{t+1}-w\, m_{\mu}^t\Big)^2\right) \ .
\end{equation}
The conditional probability of $\mathbf{m}^t$ given the hidden unit $\mathbf{h}^t$ can be easily written to the leading order in $N$, with the result
\begin{eqnarray}
P\big( m_{\mu}^{t}|h_{\mu}^t \big) &=&\exp{\left(-\sum \limits_{i=1}^N  \log 2\cosh\bigg(w\,\sum \limits_{\mu=1}^M \xi_i^{\mu} h_{\mu}^t\bigg)  \right)} \\
&\times& \exp{\left( N\bigg(w\,\sum \limits_{\mu=1}^M h_{\mu}^t\,m_{\mu}^t + \mathcal{S}^{\text{Hop}}(\mathbf{m}^t)\bigg)\right) } \ . \nonumber 
\end{eqnarray} 
 
Similarly to the CW case, the probability of going from one minimum of the free energy landscape to another in $T$ steps of AGS can be expressed as
\begin{equation} 
 P\left(\mathbf{m}^T|\mathbf{m}^0\right) =  \exp{\left(- N \min\limits_{\{m^t,h^t\}}\Phi\left(\{\mathbf{m}^t,\mathbf{h}^t\}\right)\right)}\ ,
\label{eq:integral_dynamics_Hop}
\end{equation}
where the action $\Phi\left(\{\mathbf{m}^t,\mathbf{h}^t\}\right)$ is the sum of
\begin{eqnarray} 
 \delta \Phi\left(t\to t+1\right) &=& \frac{1}{2} \sum _\mu (h_{\mu}^{t+1}-w\, m_{\mu}^t)^2 \ \\
 &+& \frac{1}{N} \sum _i \log2 \cosh\left(w\,\sum _\mu \xi_i^{\mu} h_{\mu}^{t+1}\right) \nonumber\\
 &-& w\sum _\mu  m_{\mu}^{t+1}\, h_{\mu}^{t+1} -   \mathcal{S}^{\text{Hop}}(\mathbf{m}^{t+1})  \nonumber \ .
\end{eqnarray}

The set of magnetizations $\mathbf{m}^t$ and hidden-unit values $\mathbf{h}^t$ minimizing the action $\Phi$ define the most likely path interpolating between two states in $T$ AGS steps. They are solutions of the following extremization equations for $\Phi$, which must be fulfilled at all steps $1\le t\le T-1$:
\begin{eqnarray}
(\lambda^*)^{t}_\mu &=&  w (h_{\mu}^{t} +h_{\mu}^{t+1})\!-\!w^2 m_{\mu}^{t} \ , \\
w(m_{\mu}^{t+1} + m_{\mu}^{t})\ & =& h_{\mu}^{t+1} +\frac{w}{N} \sum _i \xi_i^{\mu} \tanh\left(w \sum _{\nu} \xi_i^{\nu} h_{\nu}^{t+1}\right) 
 \ .\nonumber
\end{eqnarray}

\subsubsection{Analytical expressions of the optimal trajectories in the \texorpdfstring{$T\to\infty$}{T to infinity} limit}

As for the CW model, we find

\paragraph{Instanton-like trajectories}, defined by
\begin{eqnarray}
h_{\mu}^{t+1} &=& w\, m_{\mu}^{t+1} =   \frac{1}{w}( \lambda^*)^{t}_\mu \ , \\
 m_{\mu}^{t} &=& \frac{1}{N} \sum \limits_{i=1}^N \xi_i^{\mu} \tanh\left(w^2 \sum \limits_{\mu=1}^M \xi_i^{\mu} m_{\mu}^{t+1}\right)  \ .  \nonumber
\end{eqnarray}
The contribution to the action associated with this AGS step reads
\begin{equation}
\delta \Phi = f\big( \mathbf{m}^{t+1}\big) -f\big( \mathbf{m}^t\big) \ . 
\end{equation}

\paragraph{Thermalization-like trajectories}, corresponding to
\begin{eqnarray}
h_{\mu}^{t+1} &=& w\, m_{\mu}^{t} =  \frac{1}{w} (\lambda^*)^{t+1}_\mu \ , \\
m_{\mu}^{t+1} &=& \frac{1}{N} \sum \limits_{i=1}^N \xi_i^{\mu} \tanh\left( w^2 \sum \limits_{\mu=1}^M \xi_i^{\mu} m_{\mu}^{t}\right) \ .  \nonumber 
\end{eqnarray}
The contribution to the action associated with such an AGS step vanishes:
\begin{equation}
\delta \Phi = 0 \ . 
\end{equation}

\paragraph{Orthogonal transformation of the weight matrix.}

The computation can be repeated for a weight matrix $\mathbf{\tilde{W}} = \mathbf{W}\times \mathbf{O}$ where $\mathbf{O}$ is an orthogonal matrix. In the limit $T \to \infty$, instanton-like and themalization-like trajectories are found, and contributions to the action for both trajectories are the same as for $\mathbf{W}$. Therefore, the barriers are identical for all rotations $\mathbf{O}$. However, contrary to the previous case where the hidden unit $h_{\mu}$ codes for the magnetization $m_{\mu}$ only ($h_{\mu} = w\, m_{\mu}$), under an orthogonal transformation of the weight matrix, the hidden unit $h_{\mu}$ represents a superposition: $h_{\mu} = w \sum \limits_{\nu=1}^M O_{\nu\mu} m_{\nu}$.

\subsubsection{Transition paths between Mattis states}

In the thermodynamic limit, the $\boldsymbol{\xi}^{\mu}$ are orthogonal. The free energy landscape $f(\mathbf{m})$ (Eq.~(\ref{eq:fm_hopfield})) exhibit a large variety of critical points when $w^{2} > 1$ \cite{amit_spin-glass_1985,amit_modeling_1989}, defined through Eq.~(\ref{defvi4}), with

\begin{equation}
\langle v_i\rangle  =  \tanh\left( w^2 \sum \limits_{\mu=1}^M \xi^\mu_i  m_\mu\right) \ .
\end{equation}
Global minima of Eq.~(\ref{eq:fm_hopfield}) are reached for magnetization with only one non zero component, called Mattis states \cite{procesi_metastable_1990}. 
Numerical experiments for finite N exhibit transitions between the Mattis states, for all orthogonal transformation $\mathbf{\tilde{W}} = \mathbf{W}\times \mathbf{O}$ (Figs.~\ref{fig:transition_Hopfield}(a) \& (c)). However, the hidden representations of the path between Mattis states may be easy or difficult to interpret depending on the orthogonal transformation (Figs.~\ref{fig:transition_Hopfield}(b) \& (d)).

Furthermore, as for CW, for large $T$ and $N$ (with $T \ll N$), the probability to go from one Mattis state to another scale as $\exp{(-N \Delta f)}$. The barrier $\Delta f$ depends on $w$ and is always positive for $w^{2} > 1$ \cite{amit_spin-glass_1985}. Therefore AGS is as inefficient as MH for sampling the Hopfield model.

\subsection{Case of structured multi-modal distributions}\label{sec:MF_models}

We now turn to a more complex case of multi-modal distributions, in which the free energy minima do not correspond to orthogonal pockets of configurations in the visible space but are structured. In addition, contrary to the previous models, the hidden units $h_{\mu}$, which can be discrete or continuous, are now subject to an arbitrary, not necessarily quadratic potential $\mathcal{U}_{\mu}(h_{\mu})$. Common potentials in the machine learning community are Bernoulli or ReLU potentials \cite{nair_rectified_2010,tubiana_emergence_2017}, see Appendix~\ref{ap:MF_models_rescaling}.

\begin{figure}[!htb]
\includegraphics{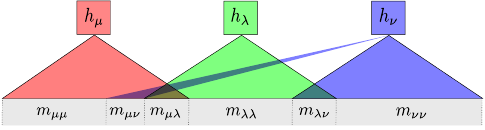}
\caption{Illustration of the structured model for $M = 3$ hidden units. The structural overlap matrix $\boldsymbol{\alpha}$ divides the visible layer into six different areas labeled by $\mu,\nu$, with $1\le\mu\le\nu\le M$. For each area, we define the corresponding normalized magnetization $m_{\mu\nu}$.} 
\label{fig:model}
\end{figure} 

The $N \to \infty$ visible units $v_i$ are $\pm \, 1 $ variables, and no potential acts on them (${\cal V}_i=0$). A visible unit $v_i$ is connected to one or two hidden units with equal weights $\frac w{\sqrt N}$, following a pattern of connections shown in Fig.~\ref{fig:model}. We define the adjacency matrix $\boldsymbol{a}$ of our model as:
\begin{equation}
a_{i\mu}=\left\{\begin{array}{ll}
1 & \text { if } W_{i\mu} = \frac w{\sqrt N}\\
0 & \text { otherwise. }
\end{array}\right.
\end{equation}
From the adjacency matrix $\boldsymbol{a}$, we define the overlap matrix $\boldsymbol{\alpha}$ and the magnetization matrix $\boldsymbol{m}$:
\begin{eqnarray}
\alpha_{\mu\mu} &=& \frac 1N \sum \limits_{i=1}^N a_{i\mu}\prod \limits_{\nu \neq \mu}(1 - a_{i\nu}) \ , \\
\alpha_{\mu\nu} &=& \frac 1N \sum \limits_{i=1}^N a_{i\mu}a_{i\nu} \ , \\
m_{\mu \mu} &= &\frac{1}{\alpha_{\mu\mu} N} \sum \limits_{i=1}^N \langle v_i\rangle\,  a_{i\mu}\prod \limits_{\nu \neq \mu} (1 - a_{i\nu})  \ , \\
m_{\mu\nu} &=& \frac{1}{\alpha_{\mu\nu} N} \sum \limits_{i=1}^N \langle v_i \rangle\, a_{i\mu}a_{i\nu} \ .
\end{eqnarray}
In other words, there are $\alpha_{\mu\mu} N$ visible units connected only to $h_{\mu}$, and $\alpha_{\mu\nu} N$ visible units connected to both $h_{\mu}$ and $h_{\nu}$. The overlap matrix $\boldsymbol{\alpha}$ partitions the visible layer into $\frac{M(M+1)}{2}$ subsets with associated magnetizations ${\bf m}$ (Fig.~\ref{fig:model}).

It is straightforward to write down the free energy per variable $f(\mathbf{m})$ as a function of the $\frac{M(M+1)}{2}$ magnetizations, with the result
\begin{equation}
 \label{eq:landscape_F(m)2}
f(\mathbf{m}) =-\sum \limits_{\mu=1}^M \hat {\Gamma}_{\mu}\left( w \sum \limits_{\nu=1}^M \alpha_{\mu\nu}\,  m_{\mu\nu} \right) - \sum _{\nu \leq \mu} \alpha_{\mu\nu} \, \mathcal{S}(m_{\mu\nu})\ , 
\end{equation}
where $\hat{\Gamma}_{\mu}$ is the rescaled cumulative generative function associated with the hidden potential $\mathcal{U}_{\mu}$, see Eq.~(\ref{eq:energy_v_RBM}) and Appendix~\ref{ap:MF_models_rescaling}, and ${\cal S}(m)$ is the entropy associated to a single $\pm \, 1$ variable with magnetization $m$. The minima of $f(\mathbf{m})$ obey the following self-consistent equations,
\begin{eqnarray}
\label{eq:self_consistent_M}
m^*_{\mu\mu} &=& \tanh \left(w \, f_{\mu}\left(I^*_\mu\right) \right) \ , \\
m^*_{\mu\nu} &=& \frac{m^*_{\mu\mu} +m^*_{\nu\nu} }{1 + m^*_{\mu\mu} m^*_{\nu\nu}} 
\ ,  \nonumber 
\end{eqnarray}
where $I^*_{\mu} = w \sum \limits_{\nu=1}^M \alpha_{\mu\nu} m^*_{\mu\nu}$ is the input received by the hidden unit $h_{\mu}$ and $f_{\mu} = \hat{\Gamma}_{\mu}^{'}$ is the transfer function associated with hidden unit $h_{\mu}$. 

\subsubsection{Optimal sampling paths with AGS}

We may now express the conditional probabilities of the magnetization matrix $\bf m$ (of dimension $M\times M$) and of the hidden-unit value vector $\bf h$ (of dimension $M$) following what was done for the simpler models in the previous sections. We first write the conditional probability of the hidden configuration given a set of visible activities,
\begin{eqnarray}\label{eq0k_M}
    P\big(h_{\mu}^{t+1}|\mathbf{m}^t\big) &=& \frac{\exp\left(-N\big(\mathcal{U}_{\mu}(h_{\mu}^{t+1}) - h_{\mu}^{t+1}I_{\mu}^t\big)\right)}{\int dh \exp\left(-N\big(\mathcal{U}_{\mu}(h) - hI_{\mu}^t\big)\right)} \\
    &\simeq & \exp\left(-N\big(\mathcal{U}_{\mu}(h_{\mu}^{t+1}) - h_{\mu}^{t+1}I_{\mu}^t\big)\right) \nonumber \\
    &\times& \exp\left(-N\hat \Gamma_{\mu}\left( I_{\mu}^t \right) \right) \nonumber \ ,
\end{eqnarray}
where we have defined the input $I_{\mu}^{t} = w \sum \limits_{\nu=1}^M \alpha_{\mu\nu} m_{\mu\nu}^{t}$ received by the hidden unit $h_{\mu}$ given the magnetization matrix $\mathbf{m}^t$.

In turn, we write the conditional probability over magnetizations given the set of hidden-unit values (to dominant order in $N$),
\begin{widetext}
\begin{equation}\label{eq2k_M}
P\big( \mathbf{m}^{t}|\mathbf{h}^t \big) 
\simeq \exp{\left(N\Bigg( \sum \limits_{\mu=1}^M I_{\mu}^t h^t_{\mu} - \alpha_{\mu\mu} \log2 \cosh\left(w \, h^t_{\mu} \right)- \sum \limits_{\mu \leq \nu} \alpha_{\mu\nu} \log2 \cosh\left(w (h^t_{\mu} + h^t_{\nu}) \right) +  \alpha_{\mu\nu}  \, \mathcal{S}(m_{\mu\nu}^{t} )\Bigg)\right)}  \ .
\end{equation} 
\end{widetext}

The probability to go from one minimum of the free energy landscape to another in $T$ steps of AGS, $P\left(\mathbf{m}^T|\mathbf{m}^0\right)$, takes the same form as Eq.~(\ref{eq:integral_dynamics}), where the action $\Phi\left(\{\mathbf{m}^t,\mathbf{h}^t\}\right)$ is the sum of 
\begin{eqnarray}\label{eqdeltaphi_M}
 \delta \Phi\left(t\to t+1\right) &=& \sum \limits_{\mu=1}^M \mathcal{U}_{\mu}(h^{t+1}_{\mu}) + \sum \limits_{\mu=1}^M \hat \Gamma_{\mu}\left( I_{\mu}^t \right) \\
&+& \sum \limits_{\mu=1}^M \alpha_{\mu\mu} \,\log 2 \cosh\left(w \, h^{t+1}_{\mu} \right) \nonumber\\
&+& \sum _{\mu\le \nu} \alpha_{\mu\nu} \log 2 \cosh\left(w (h^{t+1}_{\mu} + h^{t+1}_{\nu}) \right) \nonumber \\
&-& \sum \limits_{\mu=1}^M (I_{\mu}^{t+1} + I_{\mu}^{t}) h^{t+1}_{\mu} \nonumber \\ 
&-& \sum _{\mu \leq \nu} \alpha_{\mu\nu}\, \mathcal{S}(m_{\mu\nu}^{t+1}) \nonumber \ .
\end{eqnarray}
Notice that the previous expression extends the model studied in Section \ref{sec:Curie_Weiss_model}, which can be recovered for $M=1$, $\alpha _{11}=1$ with a quadratic potential  ${\cal U}(h)=\frac {h^2}2$.

\begin{figure}[!htb]
\includegraphics{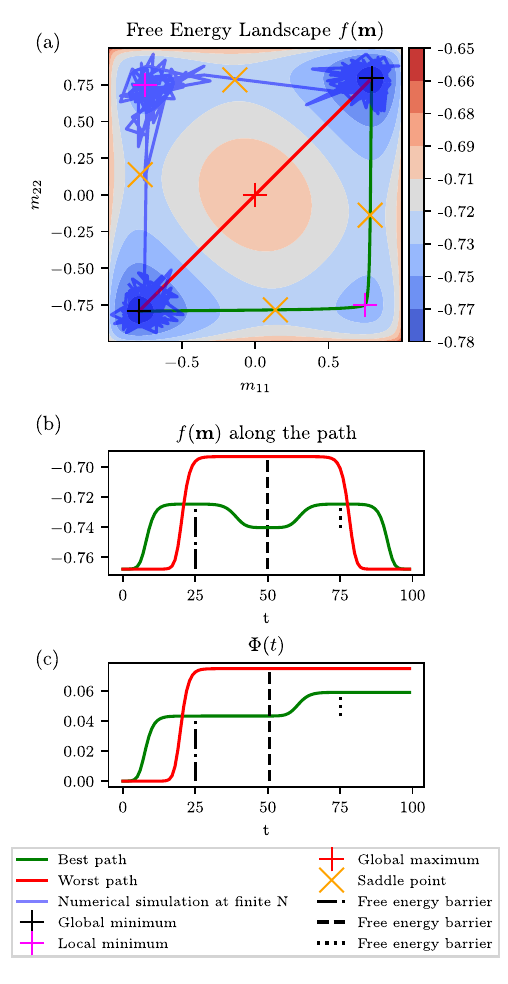}
\caption{(a) Free energy landscape for a coupled Curie-Weiss model with two global minima and two local minima. $M = 2$, $\mathcal{U}(h) = \frac{h^2}{2}$, $w=1.15 \, \sqrt 2$ and $\alpha_{12} = 0.02$. Among the many paths connecting the two global minima in $T=100$ steps, the green path is the optimal one. The red path is another path, along which both magnetizations $m_{11}$ and $m_{22}$ are equal at all times. The blue path is a representative trajectory found by simulating AGS for $N = 400$ and $10^5$ steps. (b) Free energy $f(\mathbf{m}^t)$ along the different paths. (c) Cost $\Phi\left(\{\mathbf{m}^t,\mathbf{h}^t\}\right)$ for the different paths.}
\label{fig:Curie-Weiss_2D}
\end{figure}

We show the best path found through minimization of $\Phi$ in the case of $M=2$ hidden units, quadratic ${\cal U}(h)$, $w>1$, and small positive overlap $\alpha_{12}$. The free energy landscape $f(\mathbf{m})$ represents two coupled Curie-Weiss models (Fig.~\ref{fig:Curie-Weiss_2D}(a)), and displays two global minima and two local minima. The green trajectory shows the most likely path connecting the two global minima in $T=100$ steps. Along this path, $m_{11}^t$ and $m_{22}^t$, and therefore $h_1^t$ and $h_2^t$, have asymmetric behaviors. In contradistinction, trajectories along which $m_{11}^t$ and $m_{22}^t$ are equal have exponentially smaller probabilities, see the red path. We elucidate this behavior below.

\subsubsection{Optimal trajectories in the \texorpdfstring{$T\to\infty$}{T to infinity} limit}

The set of magnetizations $\mathbf{m}^t$ and hidden-unit values $\mathbf{h}^t$ minimizing the action $\Phi$ define the most likely path, with AGS, capable of moving the system from one state to another in $T$ alternating sampling steps. They are solutions of the following extremization equations for $\Phi$, which must be fulfilled at all steps $1\le t\le T-1$:
\begin{eqnarray}\label{eqmotion2}
I_{\mu}^t + I_{\mu}^{t+1} &=& \mathcal{U_{\mu}}^{'}(h_{\mu}^{t+1})  + w\, \alpha_{\mu\mu}\, \tanh (w h_{\mu}^{t+1} ) 
\\ &+& w \sum \limits_{\nu \neq \mu} \alpha_{\mu\nu} \tanh \left( w \left( h_{\mu}^{t+1} + h_{\nu}^{t+1} \right) \right) \ ,\nonumber \\
m_{\mu\mu}^t &=&\tanh\left( w \left(h_{\mu}^{t+1} +h_{\mu}^t \right) - w \,\hat \Gamma_{\mu}^{'} \left(I_{\mu}^t \right) \right) \ ,\\
m_{\mu\nu}^t &=&   \frac{m_{\mu\mu}^t +m_{\nu\nu}^t}{1+ m_{\mu\mu}^t m_{\nu\nu}^t}\ .
\end{eqnarray}
In the infinite $T$ limit, these equations of motion admit two distinct solutions.

\paragraph{Instanton-like solutions} correspond to an increase of free energy from a local minimum, to a saddle-point of $f({\bf m})$. These solutions can be written as
\begin{eqnarray}
  \label{eq:instantonic_MF_1}
  h_{\mu}^{t+1} &=& f_{\mu}(I_{\mu}^{t+1}) \ ,  \\
   \label{eq:instantonic_MF_2} \,
  m_{\mu\mu}^{t} &=& \tanh (w f_{\mu}(I_{\mu}^{t+1}))  \ .  \nonumber 
\end{eqnarray}
Inserting these equations into Eq.~(\ref{eqdeltaphi_M}):
\begin{equation}
    \delta \Phi = f(\mathbf{m}^{t+1})-f(\mathbf{m}^{t}) \ .
\end{equation}

\paragraph{Thermalization-like solution} makes the free energy decrease until a local minimum is reached.
The relaxation solution can be written as:
\begin{eqnarray}
  \label{eq:relaxation_MF_1}
  h_{\mu}^{t+1} &=& f_{\mu}(I_{\mu}^{t})  \ ,  \\
  \label{eq:relaxation_MF_2}
  m_{\mu\mu}^{t+1} &=& \tanh (w \, f_{\mu}(I_{\mu}^{t}))  \ .  \nonumber
\end{eqnarray}
Inserting these equations into Eq.~(\ref{eqdeltaphi_M}):
\begin{equation}
    \delta \Phi = 0 \ .
\end{equation}

While instantonic and thermalization trajectories are, strictly speaking, defined for $T\to\infty$ qualitatively analogous bouts of trajectories are observed for finite $T$, see Fig.~\ref{fig:Curie-Weiss_2D}(b) \& (c) for the $M=2$ example above. The green and the red paths are each composed of a sequence of instantonic and thermalization stretches. In the case of the red path, starting from a global minimum, the instantonic dynamics leads to the global maximum of $f(\mathbf{m})$. The relaxation dynamics then brings the system down to the other global minimum. In the case of the green path, starting from a global minimum, the instantonic solution leads to a saddle point of $f(\mathbf{m})$, which is unstable for the instantonic and the thermalization dynamics. Then, the relaxation dynamics leads to a local minimum of $f(\mathbf{m})$. Through another pair of instantonic/relaxation dynamics, the second global minimum is finally reached. Thus, for the green and the red paths, the action $\Phi\left(\{\mathbf{m}^t,\mathbf{h}^t\}\right)$ corresponds to the sum of the free energy barriers along the paths (Figs.~\ref{fig:Curie-Weiss_2D}(b) \& (c)). These theoretical findings are corroborated by running AGS on a RBM with $N=400$ spins, with the same overlap matrix $\boldsymbol{\alpha}$. Along the transition path allowing the RBM to interpolate from one global state to the other, hidden units are preferentially flipped one by one, see the blue path in Fig.~\ref{fig:Curie-Weiss_2D}(a).

\subsubsection{Dependence of barrier upon structural overlap \texorpdfstring{$\alpha$}{}}

This section examines the influence of the structural overlap on the free energy barrier (and on the transition time) separating states. For the sake of simplicity, we focus on the case of $M =2$ hidden units subject to quadratic potentials and restrict ourselves to small overlap values, $\alpha = \alpha_{12} \ll 1$. For $\alpha = 0$ the two global minima of $f(\mathbf{m})$ are $ \mathbf{m}^{*}$ and $- \mathbf{m}^{*}$, where
 \begin{equation}
     \mathbf{m}^{*}  = \begin{bmatrix}
          m_{11} = m^* \\
          m_{22} = m^* 
         \end{bmatrix}.
 \end{equation}
An optimal path between these two global minima follows the sequence of critical points:
 \begin{equation}\label{transitionp0al}
    \begin{bmatrix}
          m^* \\
          m^* 
         \end{bmatrix}
         \rightarrow
          \begin{bmatrix}
          0 \\
          m^* 
         \end{bmatrix}
         \rightarrow
          \begin{bmatrix}
          -m^* \\
          m^* 
         \end{bmatrix}
         \rightarrow
          \begin{bmatrix}
          -m^* \\
          0
         \end{bmatrix}
         \rightarrow
          \begin{bmatrix}
          -m^* \\
          -m^*
         \end{bmatrix} \ ,
 \end{equation}
and, for large $T$, $\Phi(T)$ equals the sum of the free energy barriers along the path
 \begin{eqnarray}
     \Phi &=&   -f\left(\begin{bmatrix}
           m^* \\
           m^* \\
         \end{bmatrix}\right) +  2 f\left(\begin{bmatrix}
           0 \\
           m^* \\
         \end{bmatrix}\right)
         - f\left(\begin{bmatrix}
           -m^* \\
           m^* \\ 
         \end{bmatrix}\right) \nonumber  \\ &=& - \log 2 + \frac{w^2}{2} (m^*)^2 + \mathcal{S}(m^*) \ .
 \end{eqnarray}
Assume now we make small changes to the weight and overlap values, {\em i.e.} $w \rightarrow w + d w, \alpha \rightarrow d \alpha$. We denote the displacement of the critical points of $f(\mathbf{m})$ by $d \mathbf{m}$, and the variations of the free energy by $d f(\mathbf{m})$. We will consider only contributions to the first order in $d \alpha$ and $d w$, 
\begin{eqnarray}
 d \mathbf{m} &=& \mathbf{m}^w d w + \mathbf{m}^{\alpha} d \alpha \ ,  \\
 d f(\mathbf{m}) &=& f^w(\mathbf{m}) d w + f^{\alpha}(\mathbf{m}) d \alpha \ . 
\end{eqnarray}
Expressions for $\mathbf{m}^w$, $\mathbf{m}^{\alpha}$, $f^w(\mathbf{m})$
and $f^{\alpha}(\mathbf{m})$ are given in Appendix~\ref{ap:perturbation}.

As the variation of $\alpha$ changes the critical points of $f(\mathbf{m})$, we have to change $w$ in order to keep fixed the two global minima $\pm \, \mathbf{m}^*$ of $f(\mathbf{m})$. Therefore, the variation of the cost $\Phi$ between an optimal path for $\alpha = d \alpha$ and one for $\alpha = 0$ defined in Eq.~(\ref{transitionp0al}) reads
\begin{equation}
    \label{eq:Delta_Phi_first}
    d \Phi =  -df\left(\begin{bmatrix}
           m^* \\
           m^* \\
         \end{bmatrix}\right) +  2\, df\left(\begin{bmatrix}
           0 \\
           m^* \\
         \end{bmatrix}\right)
         - df\left(\begin{bmatrix}
           -m^* \\
           m^* \\
         \end{bmatrix}\right) \ .
\end{equation}

\begin{figure}[!htb]
\includegraphics{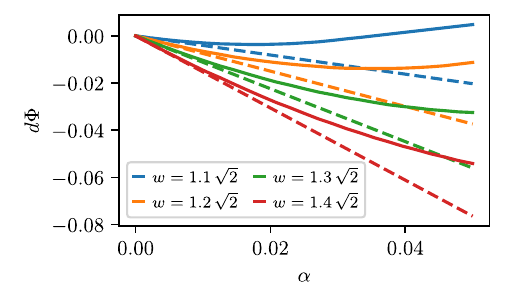}
\caption{Solid lines: numerical evaluation of $d\Phi$. Dashed lines: first order perturbation theory evaluated with Eq.~(\ref{eq:Delta_Phi_first}).}
\label{fig:perturbation}
\end{figure}

As we observe in Fig.~\ref{fig:perturbation}, a small overlap $\alpha$ reduces the cost for a wide range of $w$ and therefore helps reduce the transition time between the global minima of $f$.

\subsubsection{Time ordering of hidden-unit changes on sampling path}

\begin{figure}[!htb]
\includegraphics{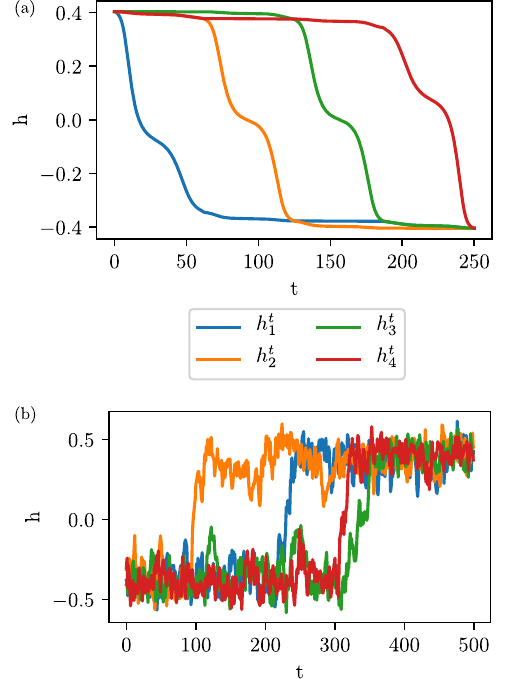}
\caption{Sampling paths for structured states. $M = 4$ hidden units are arranged on a ring, with $w = 2.2$ and $\alpha = 0.02$. (a) Numerical minimization of $\Phi\left(\{\mathbf{m}^t,\mathbf{h}^t\}\right)$ for $T = 250$. Hidden units are flipped according to their ordering on the ring $(h_1 \rightarrow h_2 \rightarrow h_3 \rightarrow h_4)$. There are $2M$ equivalent optimal paths. (b) Numerical experiment on a RBM with $N = 400$ visible units. Hidden units are flipped according to their ordering on the ring $(h_2 \rightarrow h_1 \rightarrow h_4 \rightarrow h_3)$.}
\label{fig:hidden_units_circle}
\end{figure}

As the optimal paths for the Alternating Gibbs Sampling are the ones that minimize the sum of the free energy barriers along the paths, the optimal paths depend strongly on the overlap matrix between the hidden units. If we impose a 1d structure with periodic boundary conditions for the overlap matrix, i.e $\alpha_{\mu\nu} = \alpha$ for $\nu = \mu - 1$ and $\nu = \mu + 1, \alpha_{\mu\mu} = \frac{1}{M}-\frac{M-1}{2}\alpha > 0$ (the hidden units are on a circle and have an overlap only with their two neighbors), the optimal path corresponds to an asymmetric behavior of the hidden units: they evolve one by one, according to their orders on the circle ($h_{\mu}$ evolves then $h_{\mu+1}$ then $h_{\mu+2}$ ...), see Fig.~\ref{fig:hidden_units_circle}.

\subsection{Numerical experiments}

We train RBM with the datasets defined in Section~\ref{sec:dataset}, then test the performances of Alternating Gibbs Sampling. The different RBM can generate high-quality configurations, but the dynamics associated with Gibbs sampling struggles to mix efficiently between the data modes.

\subsubsection{BAS}
\begin{figure}[!htb]
\includegraphics{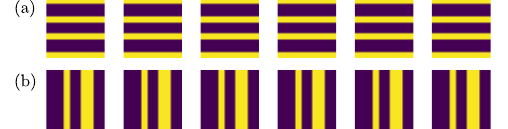}
\caption{Example of configurations obtain with AGS starting from a stripe (a) or a bar (b). 1000 steps between each frame.}
\label{fig:dynamics_AGS_BAS}
\end{figure}

We train RBM with $2L$ real hidden units subject to quadratic potentials and $\pm \, 1$ visible units. A $L_1$ regularization is added to the log-likelihood to enforce the sparsity of the weights. With this regularization, each hidden unit focuses on a given bar or a given stripe, see Section~\ref{sec:basmore} for further details. Hidden units identify the relevant degrees of freedom of the visible units. For an image of bars, hidden units encoding the bars are strongly magnetized, and the hidden units encoding the stripes are weakly magnetized (they are silent). It is essential to use real hidden units because each hidden unit must have more than two equilibrium positions (strongly magnetized with positive or negative value, and weakly magnetized with positive or negative value). This behavior is not possible with discrete units like Bernoulli or Spin.
AGS is inefficient for large $L$ and long training, and the dynamics gets stuck in a bar or stripe configuration (Fig.~\ref{fig:dynamics_AGS_BAS}). For short training, dynamics can escape from a given configuration but sampled configurations are noisy.

\subsubsection{MNIST 0/1}
\begin{figure}[!htb]
\includegraphics{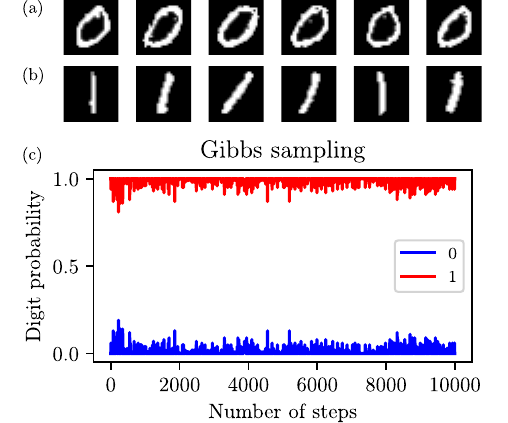}
\caption{Examples of digits obtained with AGS starting from a 0 (a) and from a 1 (b); 1000 steps between each frame. (c) Probabilities that the visible unit configurations sampled by the RBM at different times are 0 (blue) or 1 (red), estimated by a random forest classifier trained on 0-1 data \cite{ho_random_1995,breiman_random_2001}. The dynamics is stuck in a given mode.}
\label{fig:dynamics_AGS_MNIST_0_1}
\end{figure}

We train Spin-Spin RBM (hidden and visible units are $\pm \, 1$ spins). The weights of the RBM encode the digits strokes. Zeros have many strokes in common, and so have ones.
Therefore, the hidden representations of each digit are close to each other (in terms of Hamming distance). AGS is efficient to sample within a digit class and generate high-quality data, (Figs.~\ref{fig:dynamics_AGS_MNIST_0_1}(a) \& (b)). However, hidden representations of the zeros and the ones are far away from each other. Therefore, many hidden units should be simultaneously flipped to go from one class to another, which is very unlikely with AGS: the dynamics remains confined to one digit class, see Fig.~\ref{fig:dynamics_AGS_MNIST_0_1}(c). Notice that this observation crucially depends on the restriction of MNIST to 0-1 digits done here. RBM trained on all ten digits sample much more efficiently all classes and can reach 1 from 0 or vice versa \cite{desjardins_parallel_2010,tubiana_emergence_2017}, as other digits carve interpolating paths in the energy landscape.

\subsubsection{Lattice Proteins} 

\begin{figure}[!htb]
\includegraphics{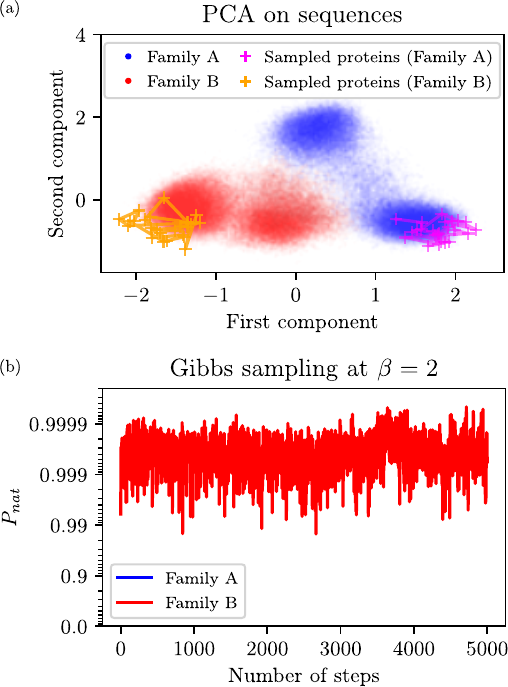}
\caption{(a) Principal Component Analysis in the sequence spaces, showing the cluster structure of each family (blue and red colors). Fuchsia and orange paths are the projection of sampled proteins with AGS, starting respectively from a protein in family A and B. Sampled proteins are stuck in a given family; 250 Gibbs steps between each cross. This number of steps is larger than the decorrelation time estimated from the Hamming distance between sequences $\mathbf{v}^t$. (b) $P_{\text{nat}}(\mathbf{v} | S)$ of sampled proteins with AGS, for $S_A$ and $S_B$, for an initial protein in the $S_B$ family (orange path in panel a). RBM generates high-quality and diverse proteins, which are different from the training data.}
\label{fig:dynamics_AGS_LP}
\end{figure}

To encode amino acids (which may take 20 values), we introduce RBM with categorical (Potts) visible units. Couplings between the hidden layer and the visible layer are represented by a $M \times N \times 20$ tensor. Thus, the energy of the RBM can be written as:
\begin{eqnarray}
E(\mathbf{v},\mathbf{h}) \!&=&\! - \sum \limits_{i=1}^N \sum \limits_{\mu=1}^M W_{i\mu}(v_i) h_{\mu}\! +\! \sum \limits_{\mu=1}^M \mathcal{U}_{\mu}(h_{\mu}) \!+ \!\sum \limits_{i=1}^N \mathcal{V}_i(v_i) \ . \nonumber
\end{eqnarray}
The weights of the RBM encode the constraints, such as contacts between different amino acids defined by the structure. Contrary to the two previous examples, the landscape has to be sampled at low temperatures to generate high-quality proteins with the RBM, i.e., proteins with a high probability to fold in a given structure, the landscape has to be sampled at low temperatures. Using the trick introduced in \cite{tubiana_learning_2019}, we copy each hidden unit $\beta \in \mathbb{N}$ times and multiply the visible fields by the same factor $\beta$:
\begin{eqnarray}
P_{\beta} (\mathbf{v}) \propto \int \prod \limits_{\mu=1}^M \prod \limits_{c=1}^{\beta} P(\mathbf{v}|h^c_{\mu}) = P(\mathbf{v})^{\beta} \ .
\end{eqnarray}
With this modification, it is possible to sample the landscape $P(\mathbf{v})$ at inverse temperature $\beta$. RBM generate high-quality proteins but struggles to mix between two families with essentially dissimilar contact maps, such as structures $S_A$ and $S_B$ defined in Fig.~\ref{fig:dataset}, see Fig.~\ref{fig:dynamics_AGS_LP}. 
Many hidden units would have to change at once, a very unlikely update with AGS to go from one family to another.

\section{Alternating Gibbs Sampling and dynamics in the latent space}
\label{sec:dynamics_hidden_space}

\subsection{Principle of the algorithm}

We have shown in the previous Section~\ref{sec:alternating_Gibbs_sampling} that AGS was as efficient as the local MH procedure to sample the landscape over the visible configurations, defined by the effective energy $E^\text{eff}(\mathbf{v})$. However, RBM offer more than this landscape, and it is natural to wonder if the representations of data could be exploited to enhance sampling performance. 
To do so, we propose a sampling algorithm combining AGS and moves in the {\em hidden unit} space, see Fig.~\ref{fig:algo}(d) and Algorithm~\ref{algo:AGS_MH_continuous}. The main idea is to exploit the fact that hidden units can encode specific features of the data. By doing Metropolis steps in the hidden space, we try to flip the hidden units one by one, or by blocks, for switching on/off the features they encode.
This flipping procedure must obviously preserve detailed balance. We therefore need to know the effective energy
over hidden configurations, $E^\text{eff}(\mathbf{h})$, defined by marginalizing the joint distribution $P({\bf v},{\bf h})$ over the visible variables
\begin{equation}
\label{eq:joint_distribution_h}
E^\text{eff}(\mathbf{h}) = - \log \left( \int d\mathbf{v}\, P(\mathbf{v},\mathbf{h})  \right)  \ .
\end{equation}

\begin{algorithm}[!htb]
\label{algo:AGS_MH_continuous}
\SetAlgoLined
 Pick $\mathbf{v}^0$ in the training set\;
 \For{$t \in \llbracket0,T\rrbracket$}{
 $\mathbf{h}^{t+1} \sim P(\mathbf{h}|\mathbf{v}^t)$\;
 $\pi$ = random permutation of $\llbracket1,M\rrbracket$\;
 \For{$i=1...M$}{$\mu=\pi(i)$\;
$h_{\mu}^{t+1} \sim P(h_{\mu}|\mathbf{h}^{t+1}_{\neg{\mu}} )$ \;
 }
 $\mathbf{v}^{t+1} \sim P\left(\mathbf{v}|\mathbf{h}^{t+1}\right)$\;
 }
 \caption{Alternating Gibbs Sampling with Metropolis-Hastings steps in latent space}
\end{algorithm}

We can gain intuition about the exponential speed up offered by the algorithm in the latent space by considering first the CW model.
In the absence of any bias (external field) between the $+$ and $-$ states of the visible variables, the effective energy $E^\text{eff}(\mathbf{h})$ is an even function of the hidden unit value $h$. A step of the sampling algorithm in the hidden space, see Algorithm ~\ref{algo:AGS_MH_continuous}, has thus probability $\frac{1}{2}$ to flip the hidden unit. Sampling back the visible layer will change the state of a macroscopic number of visible variables. Using MH algorithm in the hidden space is similar to using cluster algorithms for the visible spins \cite{swendsen_nonuniversal_1987,wolff_collective_1989}. For ferromagnetic models, these algorithms are known to be much more efficient than local MH over spins \cite{ray_mean-field_1989,persky_mean-field_1996,long_power_2014}. The latent variable is here attached to the relevant collective mode (global reversal) of the spin variables.

\begin{figure}[!htb]
\includegraphics{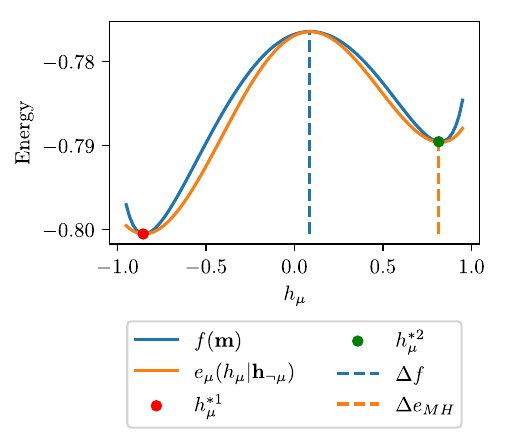}
\caption{Barriers in a structured model with $M=5$ hidden units, with $w=1.2  \,  \sqrt{5}$, $\alpha_{\mu\nu}=0.03$ for all pairs $\mu\ne \nu$.
All hidden units are frozen except $h_{\mu}$. For small overlap between the hidden units, the potential $e_{\mu}(h_{\mu}|\mathbf{h_{\neg{\mu}}})$ has two local minima for two different values of $h_{\mu}$, $h^{1*}_{\mu}$ and $h^{2*}_{\mu}$. By sampling back the visible layer $P(\mathbf{m}|\mathbf{h})$, we see that there are two local minima for $f(\mathbf{m})$. Flipping the hidden unit $h_{\mu}$ allows one to go from one local minimum to another. The free energy barrier in the hidden space with Metropolis-Hastings algorithm $\Delta e_{\text{MH}}$ is smaller than the free energy barrier of the Alternating Gibbs Sampling $\Delta f$.}
\label{fig:MH}
\end{figure}

For the mean-field structured models defined in Section~\ref{sec:MF_models}, as long as the overlap between the hidden units is weak, the hidden units could be flipped one by one for moderate system size $N$. We define the potential acting on one hidden unit, say $h_\mu$, conditional to the other units ${\bf h}_{\neg \mu}$ through
\begin{equation}
	e_\mu (h_\mu |\mathbf{h}_{\neg \mu} )= \frac{1}{N} E^\text{eff}\big(\mathbf{h}=(h_\mu, \mathbf{h}_{\neg \mu})\big) \ .
\end{equation}
Each flip of a hidden unit corresponds to a move from one local minimum to another in the landscape $e_\mu (h_\mu |\mathbf{h}_{\neg \mu} )$, see Fig.~\ref{fig:MH}. Metropolis steps in the hidden space can speed up the dynamics: the free energy barrier for Metropolis-Hastings in the hidden space, $N \Delta e_{\text{MH}}$, where
\begin{eqnarray}
\Delta e_{\text{MH}}=-\frac 1N \log \left[ \frac{\int \limits_{0}^{\infty} dh \,e^{-N 	e_\mu (h |\mathbf{h}_{\neg \mu} )}}{\int \limits_{- \infty}^0 dh \, e^{-N e_\mu (h |\mathbf{h}_{\neg \mu} )}} \right] \ ,
\end{eqnarray}
is smaller than the free energy barrier $N \Delta f$ `seen' by Alternating Gibbs Sampling.

\subsection{Application to BAS}
\label{sec:basmore}

\begin{figure}[!htb]
\includegraphics{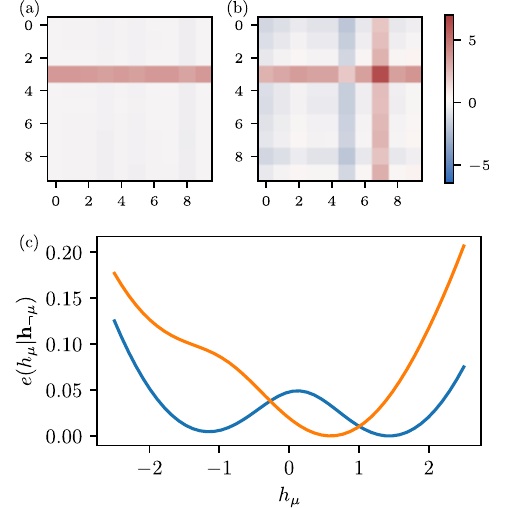}
\caption{Example of weights learned by RBM on BAS, $L = 10$. (a) With $L_1$ regularization. Each hidden unit focuses on a bar or stripe. (b) Without $L_1$ regularization. Each hidden unit focuses on several bars and stripes. (c) Potential $e_\mu(h_{\mu}|\mathbf{h}_{\neg\mu})$ 
for $\bf h$ associated with a stripe image; the minimum of the energy is set to zero. Solid blue line: hidden unit $h_\mu$ encoding a stripe; the two minima coding from the on/off stripe have roughly the same energy. Solid orange line: hidden unit $h_\mu$ encoding a bar, the minimum encoding the on bar has an energy much higher than the one corresponding to the off bar.}
\label{fig:weights_BAS}
\end{figure}

\begin{figure}[!htb]
\includegraphics{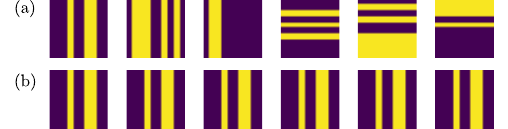}
\caption{Visible configurations obtained with Alternating Gibbs Sampling and Metropolis-Hastings algorithm in the hidden space, $L = 10$. 25 Gibbs steps between each frame. (a) With $L_1$ regularization. (b) Without $L_1$ regularization. }
\label{fig:BAS}
\end{figure}

We train RBM trained on BAS with a $L_1$ regularization to enforce the sparsity of the weights. Each hidden unit focuses on a given bar or a given stripe thanks to the regularization (Fig.~\ref{fig:weights_BAS}(a)). The change $h_{\mu} \leftarrow - h_{\mu}$ leaves the energy $E^{\text{eff}}(\mathbf{h})$ unchanged: a bar or a stripe can be present or not (Fig.~\ref{fig:weights_BAS}(c)). We use a Gibbs sampling in the hidden space where one hidden unit is updated according to Algorithm~\ref{algo:AGS_MH_continuous}.
Our algorithm efficiently switches on/off these hidden units (Fig.~\ref{fig:BAS}(a)).

Notice that, without regularization, each hidden unit would focus on several bars and stripes (Fig.~\ref{fig:weights_BAS}(b)). 
In that case, allowing for steps in the hidden-unit space does not help, and our algorithm is inefficient (Fig.~\ref{fig:BAS}(b)).

\subsection{Application to the Hopfield model}

\begin{figure*}[!htb]
\includegraphics{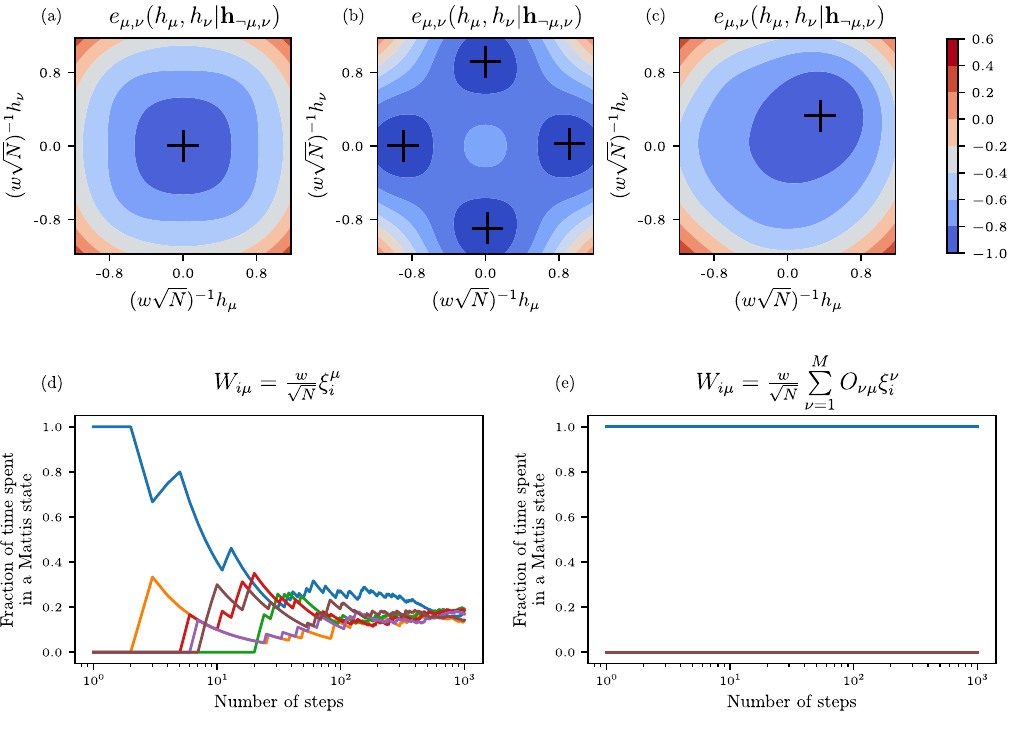}
\caption{Hopfield model encoded by a RBM with $N = 128$, $M = 6$ and $w = 1.5$ and orthogonal $\boldsymbol{\xi}^{\mu}$. (a), (b) and (c) represent the landscape $e_{\mu,\nu} (h_\mu ,h_\nu|\mathbf{h}_{\neg \mu,\nu} )$, where the $M-2$ other components of $\mathbf{h}$ are fixed. Black dots represent minima of the landscape.
(a) $W_{i\mu} = \frac{w}{\sqrt{N}} \xi^{\mu}_i$. Initial configuration is $h_{\lambda}$ strongly magnetized and $h_{\mu} \sim h_{\nu} = \mathcal{O}(1)$ . Minimum is reached for $h_{\mu} \sim h_{\nu} = \mathcal{O}(1)$. (b) $W_{i\mu} = \frac{w}{\sqrt{N}} \xi^{\mu}_i$. Initial configuration is $h_{\mu}$ strongly magnetized and $h_{\nu} = \mathcal{O}(1)$. Four minima exist corresponding to the four possible Mattis states. (c) Case $W_{i\mu} = \frac{w}{\sqrt{N}} \sum \limits_{\nu=1}^N O_{\mu\nu} \xi^{\nu}_i$. There exist only one minimum. (d) and (e): $\mathbf{v}^t$ are generated with AGS with MH steps in the hidden space, starting from 
$\boldsymbol{\xi}^{1}$. The fraction of time spent in a Mattis state is shown as a function of the number of sampling steps. (d) $W_{i\mu} = \frac{w}{\sqrt{N}} \xi^{\mu}_i$: the visible configuration $\mathbf{v}^t$ eventually visits all Mattis states with equal probabilities. (e) $W_{i\mu} = \frac{w}{\sqrt{N}} \sum \limits_{\nu=1}^N O_{\mu\nu} \xi^{\nu}_i$: the dynamics gets stuck in a given Mattis state.}
\label{fig:MH-Hopfields}
\end{figure*}

We have seen in Section~\ref{sec:hop} that, for large enough weight amplitude $w$, the AGS dynamics is stuck in one Mattis state of the Hopfield model, i.e., the magnetization $\mathbf{m}$ has only one component different from zero in the infinite size limit. The behavior of the hidden-unit configurations depends on the prescription of the weights, which may or may not be aligned with the states $\boldsymbol{\xi}^{\mu}$ (Eq.~(\ref{eq:weights_Hebb})).

\subsubsection{Aligned weights}

Let us first assume that the weights are aligned with the states, {\em, i.e.} that Eq.~(\ref{eq:weights_Hopfield}) holds. The effective energy over the hidden configurations reads
\begin{equation}
\label{eq:E(h)_Hopfield}
    E^{\text{eff}}(\mathbf{h}) = \sum \limits_{\mu} \frac{h_{\mu}^2}{2} - \sum \limits_{i} \log 2 \cosh \left(\frac{w}{\sqrt N} \sum \limits_\mu \xi^{\mu}_i h_{\mu} \right) \ .
\end{equation}
Identifying $\frac{h_\mu}{w \sqrt N} = m_\mu$, the effective energy is equal to the free energy of the Hopfield model derived in \cite{amit_spin-glass_1985} at inverse temperature $w^2$. The representations of the Mattis states are very simple in the hidden space of the RBM. In the presence of $\boldsymbol{\xi}^{\mu}$ on the visible layer, one hidden unit, say, $\mu=1$, is strongly magnetized: $h_{1} = \mathcal{O}(\sqrt{N})$. The $M-1$ other hidden units are weakly activated: $h_{\nu} = \mathcal{O}(1)$ for $\nu\ge 2$. $E^{\text{eff}}(\mathbf{h})$ has $2M$ global minima corresponding to the $2M$ Mattis states.

\paragraph{Single unit potential.} According to Eq.~(\ref{eq:E(h)_Hopfield}) the potential over the strongly magnetized hidden unit $\mu=1$ reads, after rescaling $h_1\to h_1/\sqrt N$,
\begin{equation}
\label{eq:E(h)_Hopfield1}
  e_1 (h_1 |\mathbf{h}_{\neg 1} )=   \frac{h_{1}^2}{2} - \log 2 \cosh (w h_{1} )  \ ,
\end{equation}
up to an additive constant.
This potential has two global, opposed minima for $w^2>1$. The situation is similar to the CW model studied above: MH steps in the hidden-unit space allow for efficient sampling on the states $\boldsymbol{\xi}^{1}$ and $-\boldsymbol{\xi}^{1}$. 

The potential on the other hidden units $\nu \ne 1$ is given by, up to an irrelevant additive constant and in the large-$N$ limit, after rescaling $h_\nu\to h_\nu/\sqrt N$,

\begin{equation}
\label{eq:E(h)_Hopfield2}
  e_\nu (h_\nu |\mathbf{h}_{\neg \nu} )=   \frac{h_{\nu}^2}{2} - ( 1 -m_{1} ^2)\, \bigg(\frac 1{N} \sum_i \xi_i^1\xi_i^\nu \bigg)\, h_\nu  \ .
\end{equation}
Sampling this quadratic potential allows to better explore the Mattis state around $\boldsymbol{\xi}^{1}$, but it does not help changing state.

\paragraph{Two-unit potential.} To speed up exploration of different states, we introduce the two-unit potentials
\begin{equation}
	e_{\mu,\nu} (h_\mu ,h_\nu|\mathbf{h}_{\neg \mu,\nu} )= \frac{1}{N} E^\text{eff}\big(\mathbf{h}=(h_\mu,h_\nu, \mathbf{h}_{\neg \mu,\nu})\big) \ ,
\end{equation}
where all but two hidden units are kept fixed. These potentials are plotted in Fig.~\ref{fig:MH-Hopfields}. Two typical behaviors are encountered:
\begin{itemize}
    \item $\mu,\nu$ are both different from 1. The two-unit potential $e_{\mu,\nu}$ is simply the sum of the single-unit potentials $e_\mu$ and $e_\nu$, see Eq.~(\ref{eq:E(h)_Hopfield2}). Therefore $e_{\mu,\nu}$ has only one global minimum (Fig.~\ref{fig:MH-Hopfields}(a)). Changing $h_\mu$ or $h_\nu$ does not allow for moving outside the state condensed $\boldsymbol{\xi}^{1}$.
     \item $\mu = 1$ and $\nu \neq 1$. Contrary to the previous case, $h_{1}$ is now a free parameter. Therefore, by tuning $h_{1}$ and $h_{\nu}$, four global minima of $e_{1,\nu}$ can be reached, corresponding to the cases where $h_{1}$ or $h_{\nu}$ are strongly magnetized (with positive or negative values), see Fig.~\ref{fig:MH-Hopfields}(b). We can exploit this structure by introducing a block Gibbs sampling in the hidden space, where two hidden units are updated simultaneously, see Algorithm~\ref{algo:AGS_MH_blocks}. The dynamics can now explore all the Mattis states very efficiently, see Fig.~\ref{fig:MH-Hopfields}(d).
\end{itemize}

\begin{algorithm}
\label{algo:AGS_MH_blocks}
\SetAlgoLined
 Pick $\mathbf{v}^0$ in the training set\;
 \For{$t \in \llbracket0,T\rrbracket$}{
 $\mathbf{h}^{t+1} \sim P(\mathbf{h}|\mathbf{v}^t)$\;
 $\pi$ = random pairing of $\llbracket1,M\rrbracket$, defining $M/2$ pairs of elements \;
 \For{$i \in  \llbracket1,M/2\rrbracket)$}{
$\mu, \nu = \pi(i)$ \;
$h_{\mu}^{t+1},h_{\nu}^{t+1} \sim P(h_{\mu},h_{\nu}|\mathbf{h}^{t+1}_{\neg \mu,\nu})$ \;
 }
 $\mathbf{v}^{t+1} \sim P\left(\mathbf{v}|\mathbf{h}^{t+1}\right)$\;
 }
 \caption{Alternating Gibbs Sampling with Metropolis-Hastings updates of two hidden units}
\end{algorithm}

\subsubsection{Rotated weights}

As already mentioned in Section~\ref{sec:hop}, the conditions in Eq.~(\ref{eq:weights_Hebb}) do not uniquely define the weight matrix $\mathbf{W}$. The Hopfield model energy is invariant under any transformation $\mathbf{W} \rightarrow \mathbf{W}\times \mathbf{O}$, where $\mathbf{O}$ is an orthogonal matrix. After this orthogonal transformation, the hidden representation of a Mattis state is delocalized: each component of $\mathbf{h}$ is strongly magnetized (of the order of $\sqrt N$). Single or two-unit potentials have one global minimum (Fig.~\ref{fig:MH-Hopfields}(c)). Therefore, Metropolis-steps in the hidden space do not speed up sampling (Fig.~\ref{fig:MH-Hopfields}(e)) unless all $M$ hidden units are simultaneously updated.

\begin{figure}[!htb]
\includegraphics{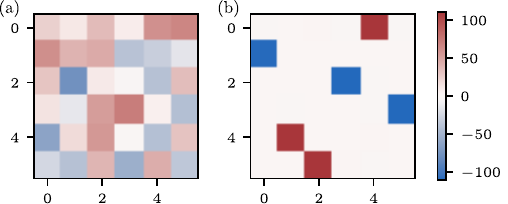}
\caption{Matrix product between the weight matrix $\mathbf{W}^{T}$ (size $M \times N$) and the matrix of patterns $\boldsymbol{\xi}$ (size $N \times M$). $N = 128$ and $M = 6$. (a) Without regularization, $\lambda_{\text{pen}} = 0$. Each pattern $\boldsymbol{\xi}^{\mu}$ has a delocalized representation in the hidden space. (b) With regularization, $\lambda_{\text{pen}} = 0.001$. Each pattern $\boldsymbol{\xi}^{\mu}$ strongly magnetized only one hidden unit.}
\label{fig:learning_Hopfield}
\end{figure}

Numerical experiments with RBM trained by gradient ascent on data sampled from the Hopfield model generally converge to a solution, where the hidden representation of a Mattis state is delocalized (Fig.~\ref{fig:learning_Hopfield}(a)) \cite{decelle_inverse_2019}. 
By adding the following penalty term in the log-likelihood, it is possible to ensure that only one hidden unit is strongly magnetized and encodes for a specific pattern $\boldsymbol{\xi}^{\mu}$, see Fig.~\ref{fig:learning_Hopfield}(b):
\begin{equation}
    LL^{\text{pen}} = - \frac{\lambda_{\text{pen}}}{L} \sum \limits_{\ell=1}^L \sum \limits_{\mu \neq \nu} |f_{\mu}(\mathbf{v}^\ell) f_{\nu}(\mathbf{v}^\ell)| \ ,
\end{equation}
where $\{\mathbf{v}^\ell\}_{\ell=1\ldots L}$ are the $L$ samples in the training set. This penalty favors solutions where only one hidden unit is strongly magnetized. Its intensity is set by the parameter $\lambda_{\text{pen}}$.

\section{Conclusion}
\label{sec:fin}

This work presents a combination of analytical and numerical results on the dynamics defined by Alternating Gibbs Sampling of Restricted Boltzmann Machines and applied to several mean-field models. We have shown how this sampling procedure can find optimal transition paths between the local minima of the free energy landscape over the visible configurations. However, large free energy barriers, extensive in the system size, have to be crossed to go from one state to another. As a result, AGS is not more efficient than standard local Metropolis sampling of the effective energy of the visible configurations. Notice that our analytical results were derived in a double large-size setting, where the asymptotics on the size $N$ of the system was considered first, and the time $T$ of transition paths was made large afterward. In practice, the probabilities that these transitions paths successfully interpolate between states are exponentially small in $N$, which implies, in turn, that transitions almost surely happen on times scales growing exponentially in $N$ (and equal to the inverse probabilities). As shown in Fig.~\ref{fig:Curie-Weiss_2D}(a), the system spends most of this exponential time attempting to escape local minima of the free energy landscapes, while transitions between the minima are actually fast (but rare). 

The inability of AGS to outperform local sampling procedures in mixing between states calls for some comments. First, it does not seem to be affected by the presence of structure in the free energy landscape. Both in the unstructured case, in which the minima of the free energy are uncorrelated (or related through global symmetries) and in the structured case, in which the minima exhibit a non-trivial organization (as observed for real data), large barriers are encountered. For structured distributions, however, the minima's non-trivial organization leads to the existence of optimal sampling paths, whose interpretation can be simpler in the hidden space of the RBM. Second, AGS, with Contrastive Divergence or Persistent Contrastive Divergence, remains an efficient training algorithm for RBM. These two procedures authorize initializations of the dynamics in different local minima close to the training data. Thus, even if AGS suffers from poor mixing between far away minima, the different minima close to the data may be well sampled. Third, AGS can be efficient when the different modes of data are connected through energy valleys. For example, AGS of RBM trained on all digits of MNIST can generate transition between 0 and 1. However, these transitions go through different intermediate states, which are other digits. When training RBM on zeros and ones only, as done in this paper, intermediate states do not exist: the two modes are not connected by low energy funnels, and transitions are unlikely to occur. Last of all, RBM are supposed to encode meaningful (hidden) representations, coding for collective features in the data. It is tempting to see these features as modes of excitation that could be flipped at once, similarly to what cluster algorithms achieve for ferromagnetic models. 

\begin{figure}[!htb]
\includegraphics{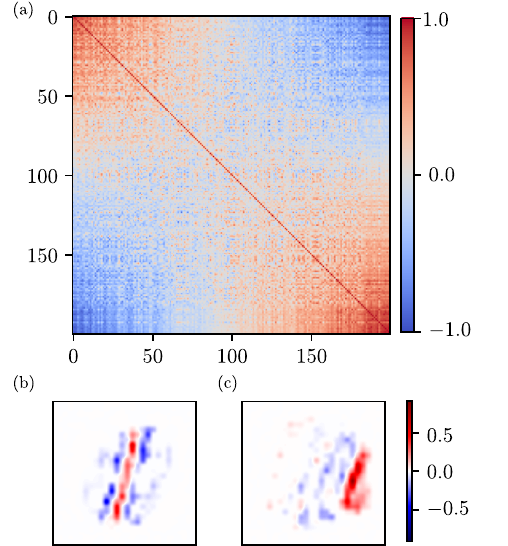}
\caption{RBM trained on MNIST 0/1, with $M =200$ hidden units. (a) Correlation matrix of the inputs received by the hidden units on the training data. Hidden units are sorted according to the components of the top eigenvector on this matrix. Two clusters emerge, corresponding to 0's and 1's: each digit is attached to roughly half the hidden units. (b) Example of weights $\mathbf{W}_{i\mu}$ for a hidden unit $\mu$ associated to 1, corresponding to a stroke specific to 1. (c) Example of weights $\mathbf{W}_{i\mu}$ for a hidden unit $\mu$ associated to 0, corresponding to a stroke specific to 0.}
\label{fig:weights_MNIST}
\end{figure}

In this context, we have shown how Metropolis-Hastings steps in the hidden space (in between the forward and backward passes of AGS) can enhance sampling performance when hidden units encode essentially independent data features or are block-correlated. Updating of one, or a small number (corresponding to the size $D$ of the block) of hidden units then allows for a macroscopic change of visible units and offers rapid mixing between states. We have illustrated this mechanism on the Bars and Stripes dataset ($D=1$) and on the Hopfield model ($D=2$). In the latter case,  the success of this procedure crucially depends on the specific set of weights output by the learning procedure, or, equivalently, on the nature of representations. MH updates in the hidden space are effective in the prototype-like regime, in which one or few strongly active hidden units rigidly determine the visible configurations \cite{cocco_statistical_2018}. This statement is expected to hold also in the so-called compositional regime, in which hidden-unit activity configurations are sparse, but the combinations of strongly activated latent variables are highly flexible and allow for a combinatorial number of visible states \cite{tubiana_emergence_2017}. In practice, one can estimate the order of magnitude of $D$ by measuring (through sampling of the trained RBM from the different data points) and clustering the covariance matrix of the hidden units.

In the case of entangled representations, in which all (or a large number of) the hidden units are strongly magnetized (with different degrees of activation from one state to another), our combined AGS-MH procedure is inefficient, as flipping a small number of hidden units is unable to change the identity of the state, and determining new, adequate configurations of a large number of hidden units would be computationally prohibitive. This phenomenon was illustrated on the Hopfield model in the case of `rotated' weights, compare Figs.~\ref{fig:MH-Hopfields}(d) \& (e). In much the same way, MH updates of a small subset of the hidden units of RBM trained on MNIST 0/1 or Lattice Proteins do not significantly enhance mixing performances. Hidden units capture features of the data, such as digit strokes for MNIST, which are correlated. Changing state demands to tune a large number of hidden units, see Fig.~\ref{fig:weights_MNIST}. In other words, the very existence of collective modes of hidden units prevents the success of our AGS-MH procedure, which is local in the hidden space. Another illustration of these collective modes in the hidden space is provided by RBM trained on BAS. Even if our algorithm is efficient to sample within a given class (bars or stripes), it cannot go from one class to another. To go from an image of bars to an image of stripes, the hidden units encoding the bars have to be silent, and the hidden units encoding the stripes have to be strongly magnetized. These define two collective modes of the hidden units, which AGS-MH cannot change. We stress that the inability of AGS to achieve rapid mixing is not limited to mean-field-like models. Even in the case of RBM tailored to encode finite-dimensional models with high-order ferromagnetic interactions, AGS suffers from poor mixing, and efficient sampling could only be obtained by combining with cluster algorithms such as the Swendsen-Wang procedure  \cite{yoshioka_transforming_2019}. In a forthcoming publication, we show how stack of RBM, with ideas proposed in \cite{bengio_better_2013,desjardins_deep_2014}, can detect collective modes of hidden units and thus improve the sampling of the energy landscape.

\begin{acknowledgments}
We are grateful to J. Tubiana for interesting discussions. Furthermore, we acknowledge fundings from Direction g\'en\'erale de l'armement (C. Roussel's PhD grant) and from the Agence Nationale de la Recherche, RBMPro Project 17-CE30-0021. 
\end{acknowledgments}

\appendix

\begin{widetext}
\section{General hidden-unit potentials}
\label{ap:MF_models}

\label{ap:MF_models_rescaling}

We consider below three different potentials acting on hidden units, and how they should scale when $N\to\infty$.

\subsection{Quadratic potential}

The quadratic potential is defined as $\mathcal{U}_{\mu}(h_{\mu}) = \frac{h_{\mu}^2}{2}$. In that case, we should rescale $h_{\mu} \rightarrow h_{\mu}/\sqrt N$. We get:

\begin{eqnarray}
P\big( h_{\mu}|\mathbf{m} \big) &=& \frac 1{\sqrt{2\pi / N}}\, \exp\left( -\frac N2 \big( h_{\mu}-I_{\mu} \big)^2\right) \ , \\
\hat \Gamma_{\mu}(I) &=& \frac{I^2}{2}  \ , \quad
f_{\mu}(I) = I \ .
\end{eqnarray}

\subsection{ReLU potential}

We can use the so-called ReLU (Rectified Linear Unit) potential $\mathcal{U}_{\mu}(h_{\mu}) = \frac{1}{2} \gamma^{+} h_{\mu}^{+2}+ \theta^{+} h_{\mu}^{+}$ where $ h_{\mu}^{+} = \max( h_{\mu},0)$, see for instance \cite{tubiana_learning_2019}. We should rescale $h_{\mu} \rightarrow h_{\mu}/\sqrt N$ and $ \theta^{+}_{\mu} \rightarrow \theta^{+}_{\mu}/\sqrt N$. We get:

\begin{eqnarray}
P\big( h_{\mu}|\mathbf{m} \big) &=& \mathcal{TN}\left(N \frac{I_{\mu}- \theta^{+}_{\mu}}{\gamma^{+}},\frac{1}{\gamma^{+}},\mathbb{R}^+\right) \ , \\
\hat \Gamma_{\mu}(I) &=& \max \left(0,\frac{1}{2}\left(\frac{I-\theta_{\mu}^{+}}{\gamma_{\mu}^{+}}\right)^2\right) \ , \quad
f_{\mu}(I) = \max \left(0,\frac{I-\theta_{\mu}^{+}}{\gamma_{\mu}^{+}}\right) \ .
\end{eqnarray}

$\mathcal{TN}(\mu,\sigma^2,\mathbb{R}^+)$ denotes the truncated Gaussian distribution of mode $\mu$, width $\sigma$ and support $\mathbb{R}^+$.
This potential is called ReLU because its transfer function is a ReLU function.

\subsection{Binary hidden units}

If the hidden units are spinlike variables, i.e. $h_{\mu} \in \{-1,1\}$, the potential can be written as a field $\mathcal{U}_{\mu}(h_{\mu}) = - c_{\mu} \, h_{\mu}$.
In that case, we should rescale $c_{\mu} \rightarrow c_{\mu} /N $, $w \rightarrow w \sqrt N $. We get
\begin{eqnarray}
P\big( h_{\mu}|\mathbf{m} \big)&=&\frac 12 \big (1 + h_\mu \tanh\left(N \left(I_{\mu}+c_{\mu}\right)\right)\big) \ ,
\\ 
\hat \Gamma_{\mu}(I) &=& |I + c_{\mu} | \ , \quad
f_{\mu}(I) = \sign \left(I + c_{\mu}\right) \ .
\end{eqnarray}

If the hidden units are Bernoulli units, i.e., $h_{\mu} \in \{0,1\}$, the potential acting on the hidden units is the same as for spins variables, and we get:

\begin{eqnarray}
P\big( h_{\mu}|\mathbf{m} \big)&=&\frac{\exp{\left(N h_\mu \big(I_{\mu}+c_{\mu}\big)\right)}}{1+\exp{\left(N\big(I_{\mu}+c_{\mu}\big)\right)}} \ ,
\\ 
\hat \Gamma_{\mu}(I) &=& \max(0,I + c_{\mu}) \ , \quad
 f_{\mu}(I) = H\left(I + c_{\mu}\right) \ .
\end{eqnarray}

$H(x)$ is the Heaviside step function. 

\section{Expansion of barrier height to first order in parameter changes}
\label{ap:perturbation}

By using first order perturbation theory with the self-consistent equation defined in Eq.~(\ref{eq:self_consistent_M}), we end up with:
\begin{eqnarray}
    \mathbf{m}^{\alpha}  = \begin{bmatrix}
           g_{\alpha} (m_{11},m_{22}) \\
           g_{\alpha} (m_{22},m_{11}) \\
         \end{bmatrix}
         \quad , \qquad
          \mathbf{m}^w = \begin{bmatrix}
           g_w(m_{11}) \\
           g_w(m_{22}) \\
         \end{bmatrix} \quad ,
\end{eqnarray}

with 
\begin{eqnarray}
 g_{\alpha}(x,y) &=& \left(-\frac{x}{2} + \frac{x + y}{1+ xy}\right)\left(\frac{2 w^2 (1-x^2)}{2-w^2(1-x^2)}\right) \ , \\
 g_w(x) &=& w x \left(\frac{2 (1-x^2)}{2-w^2(1-x^2)}\right) \ .
\end{eqnarray}

Inserting these results in the expression of $f(\mathbf{m})$ (Eq.~(\ref{eq:landscape_F(m)2})) leads to:

\begin{eqnarray}
 f^{\alpha}(\mathbf{m}) &=& -\frac{w^2}{2} m_{11} \left(\frac{m_{11}+m_{22}}{1+m_{11}m_{22}} + \frac{g_{\alpha}(m_{11},m_{22})-m_{11}}{2}  \right) -\frac{w^2}{2} m_{22} \left(\frac{m_{11}+m_{22}}{1+m_{11}m_{22}} + \frac{g_{\alpha}(m_{22},m_{11})-m_{22}}{2}  \right )  \nonumber \\ 
    &+& \frac{\mathcal{S}(m_{11})+\mathcal{S}(m_{22})}{2} -\mathcal{S}(m_{12}) +  \frac{g_{\alpha}(m_{11},m_{22})}{2} \text{arctanh} (m_{11}) + \frac{g_{\alpha}(m_{22},m_{11})}{2} \text{arctanh} (m_{22}) \quad , \\
    f^w(\mathbf{m}) &=& -\frac{w^2}{2} \left(m_{11} \frac{g_w(m_{11})}{2} + m_{22} \frac{g_w(m_{22})}{2} \right) - \frac{w}{4} \left(m_{11}^2 + m_{22}^2\right)  \nonumber \\  &+& \frac{g_w(m_{11})}{2} \text{arctanh} (m_{11}) + \frac{g_w(m_{22})}{2} \text{arctanh} (m_{22}) \quad .
\end{eqnarray}

\section{Sampling in the hidden space}
\label{ap:sampling_hidden}

Numerically, $P(h_{\mu}|\mathbf{h}_{\neg{\mu}} )$ (Algorithm~\ref{algo:AGS_MH_continuous}) and $P(h_{\mu},h_{\nu}|\mathbf{h}_{\neg \mu,\nu})$ (Algorithm~\ref{algo:AGS_MH_blocks}) are discretized and the new candidate is drawn from the discretized distribution with the tower sampling algorithm \cite{krauth_statistical_2006}.
Let us denote the acceptance probability from a configuration $\mathbf{h}$ to a configuration $\mathbf{h'}$ by $A_h(\mathbf{h} \rightarrow \mathbf{h'})$.
The Metropolis-Hastings algorithm and Gibbs sampling satisfy detailed balance in $E^{\text{eff}}(\mathbf{h})$, hence
\begin{eqnarray}
P(\mathbf{h})A_h(\mathbf{h} \rightarrow \mathbf{h'}) = P(\mathbf{h'})A_h(\mathbf{h'} \rightarrow \mathbf{h'})\ .
\end{eqnarray}

For the dynamics defined in Fig.~\ref{fig:algo}(d), we have the following acceptance probability from a configuration $\mathbf{v}$ to a configuration $\mathbf{v'}$ 

\begin{eqnarray}
A_v(\mathbf{v} \rightarrow \mathbf{v'}) =  \int d\mathbf{h}d\mathbf{h'} P(\mathbf{h}|\mathbf{v})A_h(\mathbf{h} \rightarrow \mathbf{h'}) P(\mathbf{v'}|\mathbf{h'}) \ .
\end{eqnarray}

Therefore,

\begin{eqnarray}
P(\mathbf{v})A_v(\mathbf{v} \rightarrow \mathbf{v'}) &=& \int d\mathbf{h}d\mathbf{h'} P(\mathbf{v})  P(\mathbf{h}|\mathbf{v})A_h(\mathbf{h} \rightarrow \mathbf{h'}) P(\mathbf{v'}|\mathbf{h'}) \nonumber \\
&=& \int d\mathbf{h}d\mathbf{h'} P(\mathbf{v})   \frac{P(\mathbf{v},\mathbf{h})}{P(\mathbf{v})} \frac{P(\mathbf{h'})A_h(\mathbf{h'} \rightarrow \mathbf{h})}{P(\mathbf{h})}\frac{P(\mathbf{v'},\mathbf{h'})}{P(\mathbf{h'})} \nonumber \\
&=&  \int d\mathbf{h}d\mathbf{h'}  P(\mathbf{v}|\mathbf{h}) A_h(\mathbf{h'} \rightarrow \mathbf{h}) P(\mathbf{v'},\mathbf{h'}) \nonumber \\
&=& P(\mathbf{v'})A_v(\mathbf{v'} \rightarrow \mathbf{v})  \ .
\end{eqnarray}
As a consequence, our algorithm satisfies the detailed balance condition.

\end{widetext}

\clearpage
\bibliographystyle{apsrev4-2}
\bibliography{Sampling_citations}

\end{document}